\documentclass[12pt,preprint]{aastex}
\usepackage{psfig}
\usepackage{amssymb}
\input{epsf.sty}
\shorttitle{ M/L Relationship at low masses}
\shortauthors{Hebb, L., Wyse, R.F.G., Gilmore, G.}
\received{}

\begin{document}
\title{Photometric Monitoring of Open Clusters I.  The Survey }
\author{Leslie Hebb\altaffilmark{1}, Rosemary F. G. Wyse\altaffilmark{1}, Gerard Gilmore\altaffilmark{2}}
\altaffiltext{1}{Johns Hopkins University, Dept of Physics \& Astronomy, Baltimore, MD 21218}
\altaffiltext{2}{Institute of Astronomy, Madingley Rd., Cambridge CB3 0HA, UK}
\email{leslieh@pha.jhu.edu, wyse@pha.jhu.edu, gil@ast.cam.ac.uk}

\begin{abstract}
Open clusters, which have age, abundance, and extinction information from studies of main-sequence turn off stars, are the ideal location in which to determine the mass-luminosity-radius relation for low-mass stars.
We have undertaken a photometric monitoring survey of open clusters in the Galaxy designed to detect low-mass eclipsing binary systems through variations in their relative light curves.  Our aim is to provide an improved calibration of the mass-luminosity-radius relation for low-mass stars and brown dwarfs, to test stellar structure and evolution models, and to help quantify the contribution of low-mass stars to the global mass census in the Galaxy.  

In this paper we present our survey, describing the data and outlining the analysis techniques.  We study six nearby open clusters, with a range of ages from $\sim 0.2$ to 4 Gyr and metallicities from approximately solar to -0.2~dex. We monitor a field-of-view of $> 1$~square degree per target cluster, well beyond the characteristic cluster radius, over timescales of hours, days, and months with a sampling rate optimised for the detection of eclipsing binaries with periods of hours to days.  Our survey depth is designed to detect eclipse events in a binary with a primary star of $\lesssim 0.3~M_{\sun}$.  Our data have a photometric precision of $\sim 3$ mmag at $I\approx 16$. 
\end{abstract}

\keywords{binaries:eclipsing -- open clusters and associations: individual (M35,M67,NGC~1647,NGC~6940,NGC~6633,IC~4756) --
stars:fundamental parameters -- stars:late-type -- stars:low-mass, brown dwarfs -- 
stars:planetary systems}

\section{Introduction} 
Mass is a fundamental property of a star as it is the
main parameter which determines the star's main sequence luminosity
and temperature, its lifetime on the main sequence,
and its subsequent evolution.  The stellar mass-luminosity relation is essential to
the determination of many important quantities, such as the stellar initial mass function.
Low mass stars are of particular interest: not only are they the most common stars,
their long lifetime means they can potentially `lock up' a significant
fraction of baryons and have a strong influence on galactic-scale physical processes
such as chemical evolution.  Yet in general, the masses of stars near the hydrogen burning limit remain poorly known. 
Furthermore, empirical measurements of mass, as well
as radius, are crucial to test our understanding of stellar
evolution (Lastennet et~al 2002).  Currently, observations of main
sequence stars between 1$M_{\sun}$ and 10$M_{\sun}$ are in good agreement with
theoretical stellar evolution models (Andersen 1991), but the models
deviate from the data in the optical bands for the majority of stars which are of low-mass,
M-dwarf type (Delfosse et~al 2000, Chabrier 2003).

The stellar {\it luminosity\/}
function has recently been determined with high reliability, especially
in galactic (open) and globular clusters, using both wide field surveys and the Hubble Space
Telescope (e.g. Paresce \& DeMarchi 2000; Elson et~al 1999;
Piotto \& Zoccali 1999).
Extensive field star surveys and open cluster surveys have provided luminosity functions
for low mass stellar and substellar objects 
(e.g. ~Bouvier et~al. 2003; Moraux et~al. 2003; Luhman et~al. 2003; Santiago, Gilmore \& Elson 1996; Gizis \& Reid 1999;);
{\tt HIPPARCOS\/} results provide an accurate and precise luminosity function for apparently
luminous stars near the Sun (e.g. Kovalevsky 1998).  The
stellar luminosity function down into the M-dwarf regime has been
determined in such diverse environments as the metal-poor, low-surface
brightness satellite galaxies of the Milky Way (e.g. the Ursa Minor
dwarf spheroidal galaxy; Feltzing, Gilmore \& Wyse 1999; Wyse et~al 2002)
and the metal-rich central bulge of the Milky Way (Zoccali et~al 2000).
These analyses show that the faint stellar {luminosity} function is universal in form, at all
luminosities and in all environments, once one allows for known other
effects, particularly binarity in field stars and mass segregation
and dynamical evolution in clusters (e.g.~Gilmore 2001).

However, conversion of a luminosity function to a {\it mass\/}
function remains problematic.  As emphasized by D'Antona (1998),
features such as local peaks in the luminosity
function may arise from structure in the mass-luminosity relation, and
do not necessarily reflect peaks in the mass function.  Unfortunately,
stellar mass cannot be determined directly for the majority of
stars.  Only stars where gravitational effects can be
analysed, such as multiple or binary systems, have measurable
masses, and these are used to calibrate the stellar mass-luminosity
relationship.  

Detached eclipsing binary stars provide the most accurate 
determinations (to $\sim 1\%$) of stellar mass and radius (Andersen 1991), 
and analysis of their orbits is a long established technique for
empirical calibration and testing of stellar structure models.
These systems place extremely stringent
constraints on our understanding of stellar evolution; they simultaneously provide
 measurements of the masses and radii for two stars which are (reasonably) 
assumed to have a single age and metallicity.
 
Although the number of well-studied eclipsing binaries with low-mass components is small.
Only three eclipsing binary objects with M-dwarf primaries (and
secondaries) are currently known: YYGem (Bopp~1974; Leung \& Schneider~1978),
CM Dra (Lacy~1977; Metcalfe~1996), and GJ~2069A (Delfosse et~al. 1999).
These systems were discovered serendipitously and are of
unknown age and of uncertain metallicity, two parameters that affect the
luminosity and thus the mass-luminosity relation.  
For example, a $0.3 M_{\sun}$ star with metallicity of -1 dex is 
predicted to be $\sim 1.2$ magnitudes brighter in the V-band than a solar 
metallicity star of the same mass (Baraffe et~al. 1997; 1998).
Furthermore, one of the three known eclipsing systems, CM~Dra, has
a high space velocity ($163 {\rm km s^{-1}}$) characterisitic of a Population II star, and thus is likely to have a
metallicity less than solar.  Although 
Viti et~al. (1997, 2002) find discrepancies between the optical and infrared results when 
they attempt to measure its metallicity, if CM~Dra is metal-poor, it will not fall along the
same mass-luminosity relation as solar metallicity objects.

Radial velocity measurements have also been used
to derive orbital solutions, and thus accurate
stellar masses, for resolved visual binary systems.
S\'egransan et~al. (2000) were able to derive stellar mass to 0.2-5\%
by measuring the radial velocity curve and angular separation of
a set of nearby visual binary M-dwarf systems. 
Further, interferometric techniques have been used to directly measure
the angular size, and thus radius, of several nearby low-mass stars
(Lane et~al. 2001; S\'egransan 2003).  Again however, these are field stars of unknown age and metallicity. 

In a study by Delfosse et~al. (2000), the set of all mass measurements to date of M-dwarf stars with errors
$\le 10\%$ (16 binary systems) was compared to theoretical models of Baraffe et~al. (1997; 1998) and Siess et~al. (2000).
All binary systems in the study have measured parallaxes (HIPPARCOS, ESA 1997; Yale General 
Catalog of trigonometric parallaxes, van Altena 1995; S\'egransan et~al. 2000; Soderhjelm 1999;
Forveille et~al. 1999; Benedict et~al. 2000), and their photometry was taken largely 
from the homogeneous compilation by Leggett (1992).

The study found that, in the J, H, and K infra-red bands, these data form a low-scatter mass-absolute magnitude sequence and 
that the theoretical mass-luminosity relation is an good match to the 
empirical fit to these points  
for M-dwarf stars between $\sim 0.6 M_{\sun}$ to $0.1 M_{\sun}$.  
The V-band mass-absolute magnitude data, on the other hand, exhibit scatter that is greater than the quoted errors in 
the measurements (see Figure~\ref{fig:magfit}).
Differing metallicities could be the source of this scatter, but
the theoretical models are also systematically offset from the mean data.
For example, the individual
components of the binary systems GJ 2069A and Gl 791.2 are $\sim 2$
magnitudes too faint in the V-band compared to 
the mean empirical mass-absolute magnitude relation (Delfosse et~al. 1999; Benedict et~al. 2000).
Further, the solar metallicity models of Baraffe et~al. (1998) are too luminous   
by $\sim 0.5$ magnitude at a given mass below $\sim 0.5
M_{\sun}$ (Delfosse et~al. 2000 and references therein)

It is a prediction of these same models that the 
metallicity is an important factor in determining the appropriate
optical mass-luminosity relation.
The spectra of late-K and M dwarfs are 
complex to analyse so that direct metallicity estimates are
uncertain (e.g. Gizis 1997; Gizis \& Reid 1997);
this is an unavoidable limitation in the use of
field binary systems.  Further, the models adopted have an age of 5 Gyr,
but there are no age estimates for the systems in the field.  A $0.1  M_{\sun}$
star is expected to dim by approximately one magnitude as it ages from 200 Myr to 5 Gyr (Baraffe et~al. 1998) and younger
stars are brighter still; this is a further source of uncertainty.

In order to provide a better calibration of the mass-luminosity-radius
relation, test models of stellar structure and evolution, and
determine the contribution of low-mass stars to the global mass
census in the Galaxy, we have undertaken a photometric
monitoring survey targeting K/M-dwarf stars ($8 \lesssim M_V \lesssim
13$) in open clusters with a range of metallicities and ages which is known
from the upper main sequence cluster members.  We have
selected for this study six open clusters with a range of ages
from $\sim 0.2 \, {\rm to}\, 4$ Gyr and metallicities from $-0.2$
dex to approximately solar (Table~\ref{tab:cluslis}).  We obtained observations of each
cluster over an area of greater than 1 square degree per target, well
beyond the characteristic radius defined by the brighter stars.  

Additional products of our survey are deep optical color magnitude
diagrams (V,V-I) over a wide FOV which will be used to determine spatial
variations in the mass functions.  We will investigate dynamical
effects in open clusters as a function of age, such as mass
segregation (e.g. Hurley et~al. 2001; Raboud \& Mermilliod 1998; Bonatto \& Bica 2003).
In addition, our data are sufficiently precise to provide improved determination of the 
intrinsic photometric stability of G-dwarf stars
in the clusters; this is a necessary input to ongoing and future
transit searches for Earth-like planets around solar analogs.

Considerable recent effort has been expended in complementary projects including variability
studies of brighter stars in open clusters (e.g.~Roberts, Craine \&
Giampapa 2000; Sandquist \& Shetrone 2001).  Variability studies of field stars and open clusters in search of planetary transits
(Udalski et~al. 2002,2003; Mall\'en-Ornelas et~al. 2003; Mochejska 2002, 2003; Street et~al. 2003) 
are also complementary.

In this first paper, we describe the observations and observational strategy of our survey, 
together with general data reduction and analysis techniques.  The introduction to the survey is given in \S~\ref{sec:obsst}.  
We go on to discuss the data reduction and calibration in \S~\ref{sec:dr} and the time-series photometry
techniques in \S~\ref{sec:phot}.  
Finally, in \S~\ref{sec:var}, we review the algorithms we developed  for the detection of intrinsic variability 
and the determination of the best-fit period of the variability. 

Future papers in this series will discuss the analysis of the
photometric data for low-mass stars in the individual clusters,
including candidate eclipsing K/M-dwarf binary systems; follow-up spectroscopy and high frequency 
photometry of candidate systems; the photometric stability of Sun-like G-dwarf stars;
the radial profiles of the clusters as traced by stars of different masses, 
and the relation to mass segregation; and the photometric search for variability of non-cluster field stars.

\section{Observational strategy} 
\label{sec:obsst}

Candidate low-mass eclipsing binary stars are detected through
variations in their apparent brightness as one star in the binary
passes in front of the other along the line of sight.  For an eclipse to occur, the orbital
inclination, $i$, must be close to $90^\circ$; this
constraint on $i$ has the disadvantage that the
probability of $i \sim 90^\circ$ is low. 
The periods and eclipse durations of known eclipsing M-dwarf binaries 
may be used to define an effective observing strategy.  
The three known systems have the following orbital
properties:  CM~Dra has a period, $P=1.27$ days, orbital inclination $i=89.82^\circ$, and duration $\tau \sim 1.2$ hours (Lacy 1977); YY Gem has $P=0.816$ days, $i=86.54^\circ$, and $\tau = 1.9$ hours (Leung \& Schneider 1978); and
GJ 2069A has $P= 2.77$ days, 
$i=86.7^\circ$, and $\tau \sim 2$ hours (Delfosse et~al. 1999). 

We perform an analytic calculation using simple scaling arguments 
to define an initial estimate of the relevant parameters for a binary search.
The total mass of a binary system with an M-dwarf primary
will not differ from $0.5 M_\odot$ by more than a factor of two either
way. The range of possible separations (and hence periods) will be
much larger, and thus one can safely adopt a total mass of $0.5
M_\odot$ as a reference.  For such a system, and inclination $i=90$,
one may calculate the eclipse duration from estimates of the angles
subtended by the stars and their angular orbital speed, with duration
$ \sim $ (angular size/angular speed).  From Kepler's laws the angular
frequency $\omega^2 \propto a^{-3}$, (with $a$ the separation between
the stars), while the angle subtended by the primary at the secondary
goes as $a^{-1}$, so there is only weak dependence (scaling like
$a^{1/2}$) of the eclipse duration on the separation.  For low mass
stars, stellar radius and mass follow $R/R_\odot \approx
M/M_\odot$, so that the angle, $\theta$, subtended at the
secondary by the primary ($M_1$) is typically $ \theta \sim 2.5 \times
10^{-3} {1 \over a(AU)} {M_{1} \over {0.25 M_\odot}}\, rad. $ Scaling from
the angular speed at 1AU for a 1~solar
mass system, close to 1~degree per day, implies that a representative eclipse duration is $1 \leq \tau \leq 5$ hours 
for periods from 1 day to 1 year.  This is well-matched to observational
convenience and, by construction, to the observed eclipse duration for the
few known M-dwarf systems. 

For random orientations of the orbital plane, the probability that $i$
lies in the range $i+{\rm d}i$ follows the usual cos($i$) scaling.  Adopting
the mass-radius relationship $M \propto R$ as above, for a system of
$0.5M_\odot$ at a separation
corresponding to a period of 1 day, an
eclipse requires that the inclination must be $i\geq 80.5^{\circ}$; at a
period of 100 days (Mercury's orbit), the required inclination is $i \geq
89.56^{\circ}$, and at a period equal to 1 year, $i \geq
89.82^{\circ}$. The corresponding probabilities of this alignment are
0.17, $8 \times 10^{-3}$ and $ 3 \times 10^{-3}$ respectively.  The
distribution function of expected separations and periods is
ill-determined.  However, it is clear from these small probabilities
of appropriate inclination angle that a sample of order several thousand
 M-dwarf stars must be observed to have a few eclipsing M-dwarf
systems.  The number of star-hours which must be observed to detect these systems
depends on the duty cycle of an eclipse which ranges from 0.05 at periods of 1 day to 0.002 at periods of 100 days. 
This requires a wide field, so that
one can efficiently survey a large number of stars repeatedly, in order to 
detect variability. 

\subsection{Monitoring Strategy: Monte Carlo Simulations}

An independent and more robust calculation may be derived through 
Monte Carlo simulations of an ensemble of binary systems.  We carried out the simulations
to calculate the probability of observing low-mass binaries in their
eclipse phase and their expected light variations (period, duration). 
These allow us to estimate the scope of the required survey (e.g.~number of star-hours)
 and the optimal timescales for monitoring the clusters.

The simulations require as input the distribution of orbital
parameters of binary systems and the intrinsic characteristics
of late-K and M-dwarf type stars, such as absolute magnitude and radius.
The distribution of orbital parameters of low-mass binary systems are not yet
established due to the difficulty of obtaining good statistics from the limited number of systems known.
Thus we based our simulations on distributions of binary parameters for G-dwarf systems given by Duquennoy \& Mayor (1991, hereafter DM91)
and, at the suggestion of the referee, those found by Halbwachs et~al. (2003, hereafter HMUA03) for F7 to K-dwarf spectroscopic binary systems.

DM91 carried out an analysis of a local, distance limited set of binary stars with
G-dwarf primaries and found (i) the period distribution is
approximately log-normal with a mean of 180 years; (ii) short period
binaries ($P<11$ days) tend to have circular orbits; (iii) binaries
with $11\le P {\rm (days)} \le 1000$ have a peaked eccentricity
distribution with $<e> \sim 0.3$; and (iv) for $P > 1000$ days the
eccentricity distribution, $f(e)$, tends toward $f(e)=2e$.  Further, these data
are consistent with the primary and secondary stars being drawn independently from
the same initial mass function (IMF).
  
The more recent paper by Halbwachs et al. 2003 agreed with the period and eccentricity distributions
defined in DM91, but observed a different mass-ratio, $q$, distribution than
that which would be derived if two stars were drawn independently from the same IMF.
HMUA03 found a q-distribution with a broad peak from $q\approx 0.2$ to 
$q\approx 0.7$ and a sharp peak for $q > 0.8$ (twins) in their sample of 89 spectroscopic binaries.
The results of HMUA03 suggest  that the number of $q > 0.8$ systems tends to decrease with increasing period 
and that these same systems tend to have smaller eccentricities than lower mass-ratio systems with similar periods. 
These trends are marginally significant and were not quantified in the paper, therefore we did not incorporate them into our simulations.

\subsubsection{Description of the Simulations}

In the simulations, we adopted the period and eccentricity distributions from DM91 and tested both the HMUA03 and DM91 mass-ratio distributions.
We populated a model open cluster with primary stars drawn from a multi-part power-law
IMF following Kroupa (2001) with a slope $\frac{{\rm dn}}{{\rm dm}} = -2.3$  for 
$m/M_\odot \ge 0.5$ and a slope $\frac{{\rm dn}}{{\rm dm}} = -1.3$  for stars of mass $0.08 \le m/M_\odot < 0.5$.
Although observations show the stellar multiplicity fraction is a function of primary mass (e.g. Sterzik \& Durisen 2003), 
we are interested only in
this fraction in the mass range of M-dwarf stars.  Therefore we created binary systems out of
35\% (e.g. Reid \& Gizis 1997 (35\%); Fischer \& Marcy 1992 (42\%); Marchal et~al. 2003) of M-dwarf stars in the simulated cluster between
$0.08$ and $0.7 M_\odot$.
We ran two sets of simulations, using DM91 and HMUA03 to define the mass of the secondary star based on the observational results in each paper.

In the first set of simulations, we chose independently a secondary
star from the same IMF (following DM91).  It is unknown whether brown dwarfs pair in binaries in 
the same fashion as solar-type stars thus we confined our simulations to only stellar mass systems, not brown dwarfs.   
A random orbital inclination,
{\it i}, was assigned to each binary ($i=90^\circ$ being edge-on), along with a period
drawn from the log-normal distribution with a mean of 180 years,
following DM91.  Knowing the period, the major axis distance, {\it a},
was then calculated using Kepler's Laws and an eccentricity was
adopted using the appropriate distribution according to DM91.  For systems defined to be on non-circular orbits, we also
chose a random angle between zero and $2\pi$ radians which defines an orientation for the elliptical orbit.

In the second set of simulations we picked a 
mass-ratio, $q$, for each binary from a distribution which reproduces the general trends found by HMUA03.  
We parameterized the $q$ distribution as the sum of three standard gaussians 
with mean, $<q>$, $\sigma$, and normalization, $A$, values of 
$<q>$=1.0/0.6/0.3, $\sigma$=0.01/0.01/0.02, and $A$=0.8/0.42/0.5 for the three gaussians.  
The ensemble of $q$ values combined with their corresponding primary masses define the masses of secondary stars in the simulated binary systems.
Again we assigned an inclination angle, a period, an eccentricity, and a viewing angle to each simulated binary system and calculated
the semi-major axis distance using Kepler's Laws.

In accordance with observational evidence for the ``brown dwarf desert'' (Marcy \& Butler 2000; Halbwachs et~al. 2000; Zucker \& Mazeh 2001), 
the lack of brown dwarf companions in short period ($a < 3$ AU) binary systems,
if a $q$ value chosen in the simulation produces a secondary star in the brown dwarf regime ($m < 0.08 M_\odot$) 
for a system with a period less than 1000 days, we reassign that primary object as a single star.  This skews the adopted mass-ratio
distribution for late type M-dwarfs to slightly higher $q$ values and introduces a small trend toward smaller binary 
fraction with decreasing primary mass. 
This is reasonable considering the mass-ratio distribution for low-mass M-dwarf stars is unconstrained and we reproduce the observed 
trend of decreasing binary fraction with decreasing primary mass while accounting for the observed brown dwarf desert.

\begin{figure}

\centerline{\hbox{\hspace{0.0in} 
\psfig{file=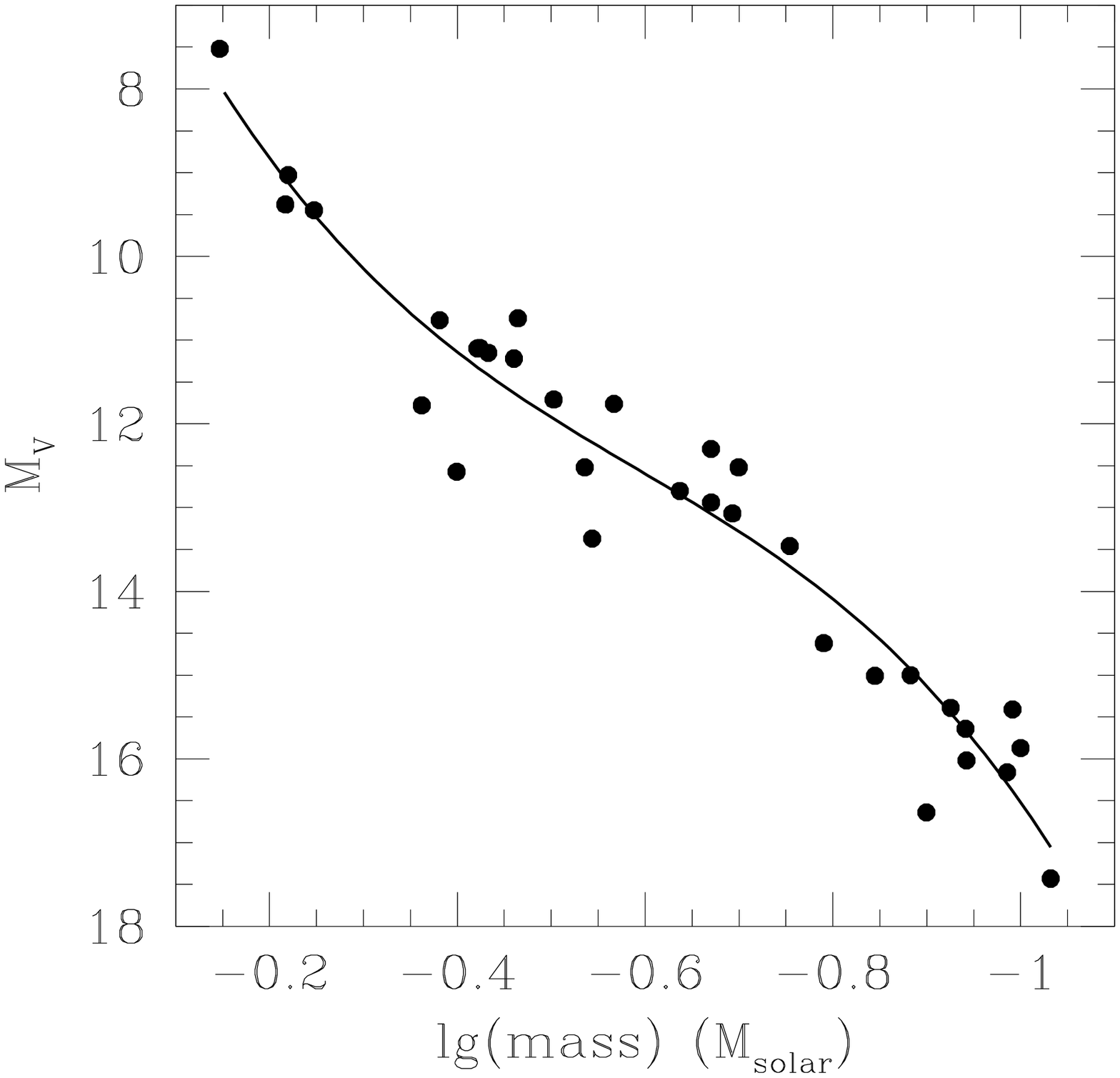,width=3.3truein}
\hspace{0.25in}
\psfig{file=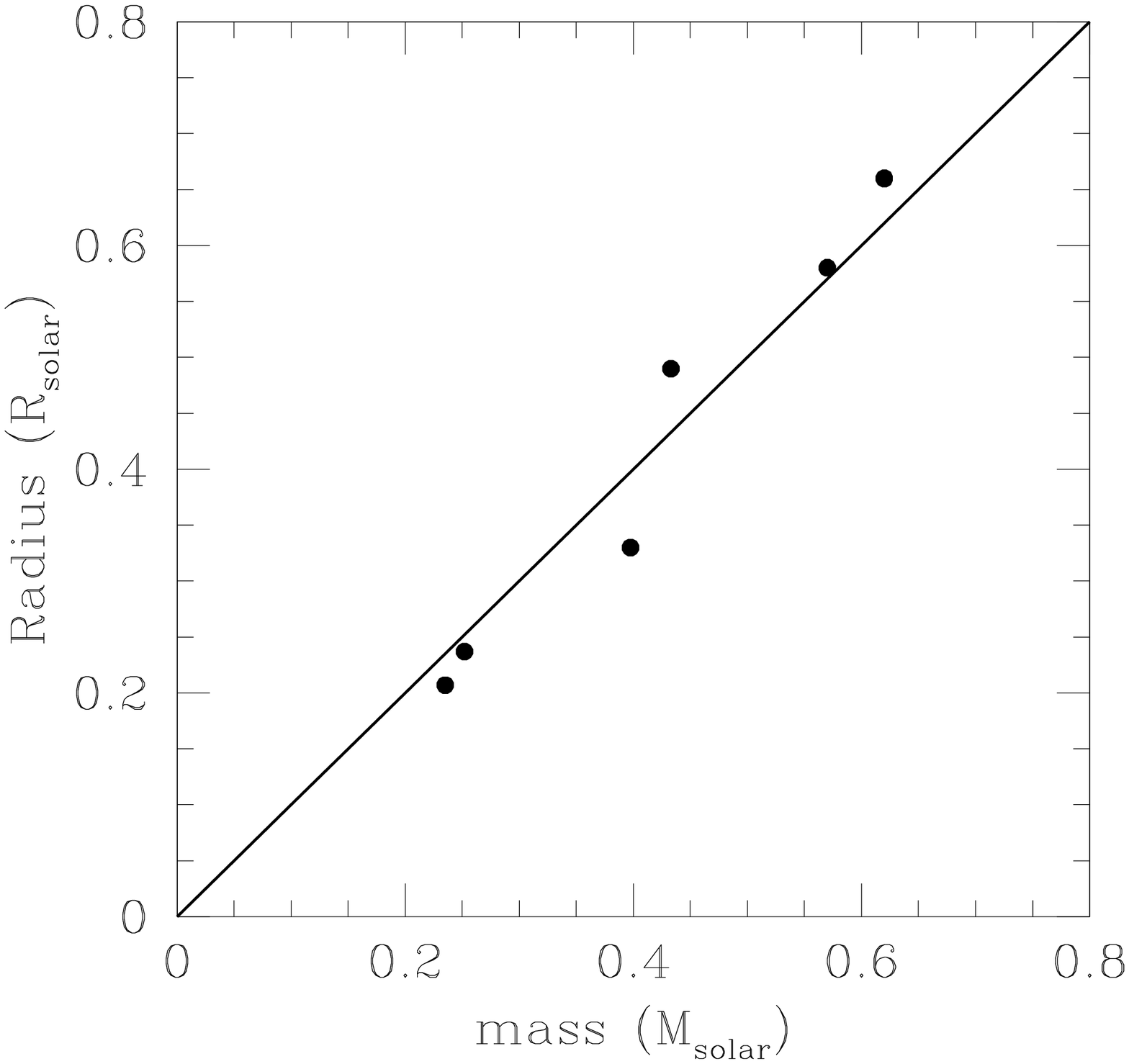,width=3.3truein}
 }
}
\vspace{10 pt}
\caption{(a) Mass-absolute V-band magnitude data compiled by Delfosse {\it et~al.} (2000) with a third order polynomial fit superposed.
 (b) Mass and radius data for the 3 known eclipsing binary M-dwarfs (YY Gem, CM Dra, and GJ 2069A) with a homology scaling 
for low-mass stars,  $\frac{\rm R}{{\rm R_\sun}} \approx \frac{m}{{\rm M_{\sun}}}$, drawn through the points. }
\label{fig:magfit}
\end{figure}

The individual components of the simulated low-mass binary systems were then assigned an absolute V-band magnitude and radius defined by
their mass.  The mass-absolute magnitude relation we adopted is based on an empirical fit to the limited existing data for M-dwarf
stars.  We fit the lowest order polynomial which reproduced the broad trends in the data compiled by Delfosse et~al. (2000).
The fit is given by 
$$M_V=4.750-26.253*log_{10}(\frac{m}{M_{\sun}})-33.168*log_{10}(\frac{m}{M_{\sun}})^2-18.684*log_{10}(\frac{m}{M_{\sun}})^3$$ 
for a given mass, $m$,
and is plotted with the data in Figure~\ref{fig:magfit}a.  As there are only three well-studied eclipsing binary M-dwarf systems 
(six stars) with accurate radius measurements, we adopted the homology scaling for low-mass stars, 
$\frac{\rm R}{{\rm R_\sun}} \approx \frac{m}{M_{\sun}}$.
We show the data together with the adopted mass-radius relation in Figure~\ref{fig:magfit}b.

Low-mass eclipsing binary systems in the simulations   
are defined as binary objects that satisfy the following criteria:  the primary component
has a mass in the range $0.08 M_{\sun} \le M_p \le 0.7 M_{\sun}$; the inclination angle allows for eclipses, requiring 
${\rm cos}(i)~<~(R_p+R_s)/r$, where $R_p$ and $R_s$ 
are the radii of the primary and secondary stars and $r$ is the shortest separation between the two binary components when viewed
at the chosen orientation;   and the system must be a detached binary, requiring that the separation, $r$,
be greater than the Roche limit of the secondary star.  

Finally, determining the likelihood of capturing the eclipsing systems while in their eclipse phase depends on the observing window
function (e.g. sampling rate, number of hours observed).  In general the short period binaries that eclipse more frequently
and spend a greater percentage of their time in the eclipse phase are most easily observed.  We applied a 
window function to the simulated low-mass eclipsing binary systems designed to reproduce a possible observing run.
We chose a starting position for the secondary star from a distribution proportional to its inverse orbital speed and
assumed observations were taken with a sampling rate of one per hour over a ten hour observing night for four consecutive nights.

\subsubsection{Results of the Simulations}

\begin{figure}
\psfig{file=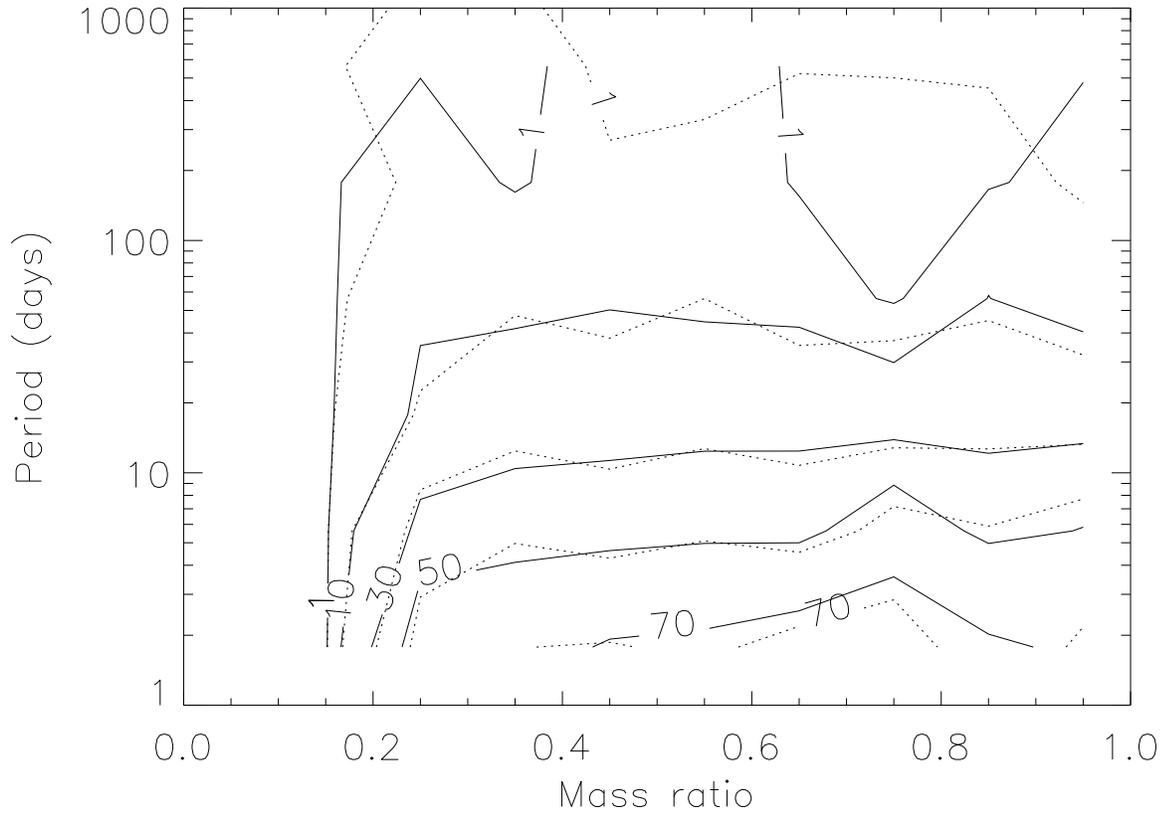,width=6truein,angle=90}
\caption{The contour plot shows the percentage of low-mass eclipsing binary systems, as a function of
mass ratio and period, which are detected in the simulations.
We assume a sampling function of one observation per hour over a ten hour observing night for 
four consecutive nights and a minimum eclipse detection amplitude of 0.05 magnitudes. }
\label{fig:qpcontour}
\end{figure}

We ran 1000 realizations of the simulation for each mass-ratio distribution and show, in Figure~\ref{fig:qpcontour},
the type of eclipsing binary systems we are likely to detect.  
The figure is a contour plot of mass ratio versus period with the contour level marking 
the percentage of low-mass ($0.2 \le m/M_{sun} \le 0.7$) eclipsing binary systems in the simulation which were detected with the sampling 
function described above and have a amplitude $\ge 0.05$ magnitudes.  The solid line is the result of the
simulations with the HMUA03 q-distribution and the dotted line is the same for the set of simulations assuming random pairing of
binaries from the same IMF (following DM91).  As expected, we are most likely to detect small period, high mass ratio systems.
Further, the two mass-ratio distributions give nearly the same results.  Although the small scale structure
of the distributions appears very different, the broad trends are not.  
$\sim 40\%$ of the binaries created using the HMUA03 q-distribution have $q>0.7$ 
and $\sim 2\%$ have $q<0.2$ as opposed 
to $\sim 30\%$ and $\sim 7\% $ if the q-distribution is derived assuming randomly paired binaries from a single IMF. 

\begin{figure}
\psfig{file=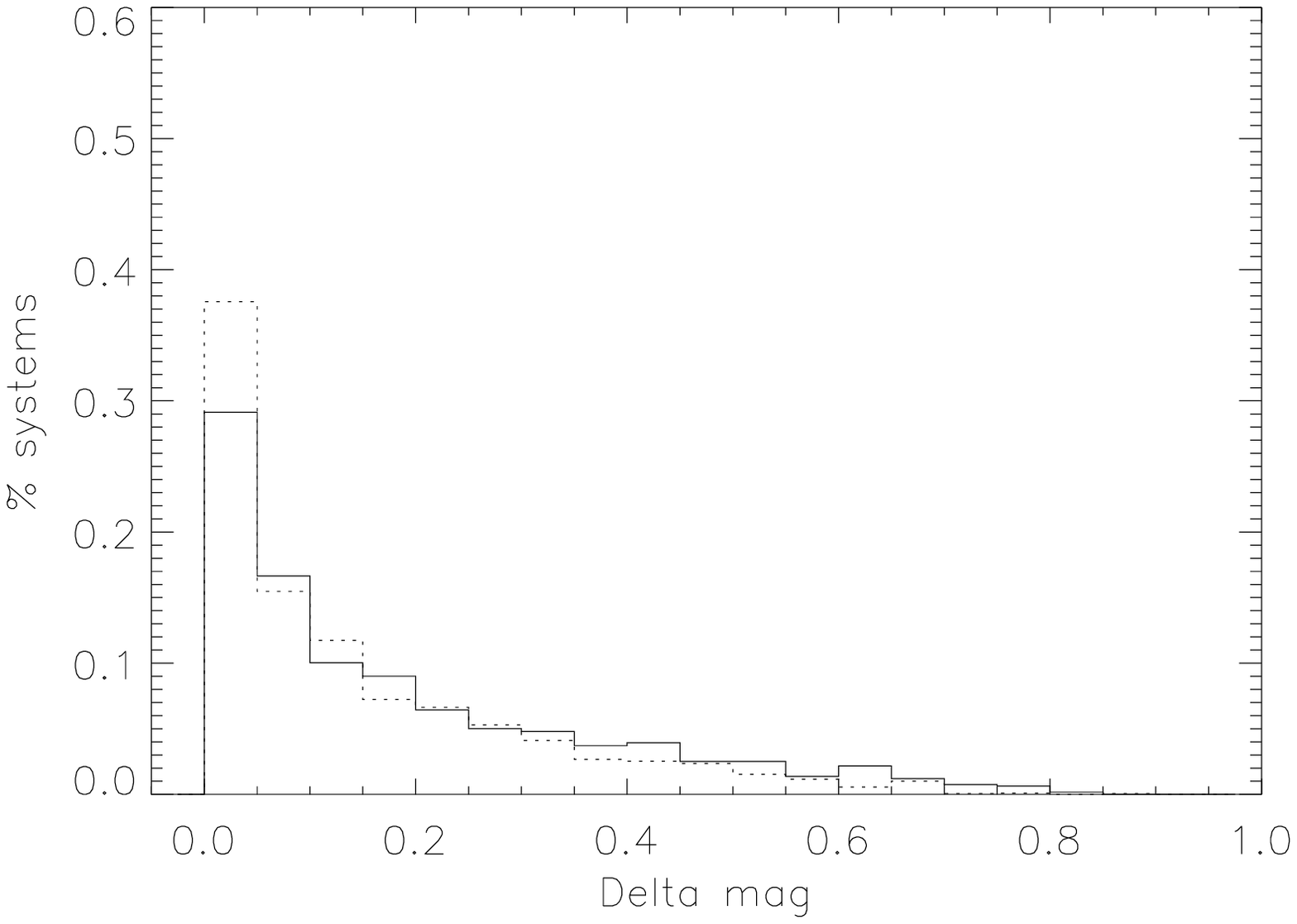,width=5truein}
\vspace{10 pt}
\psfig{file=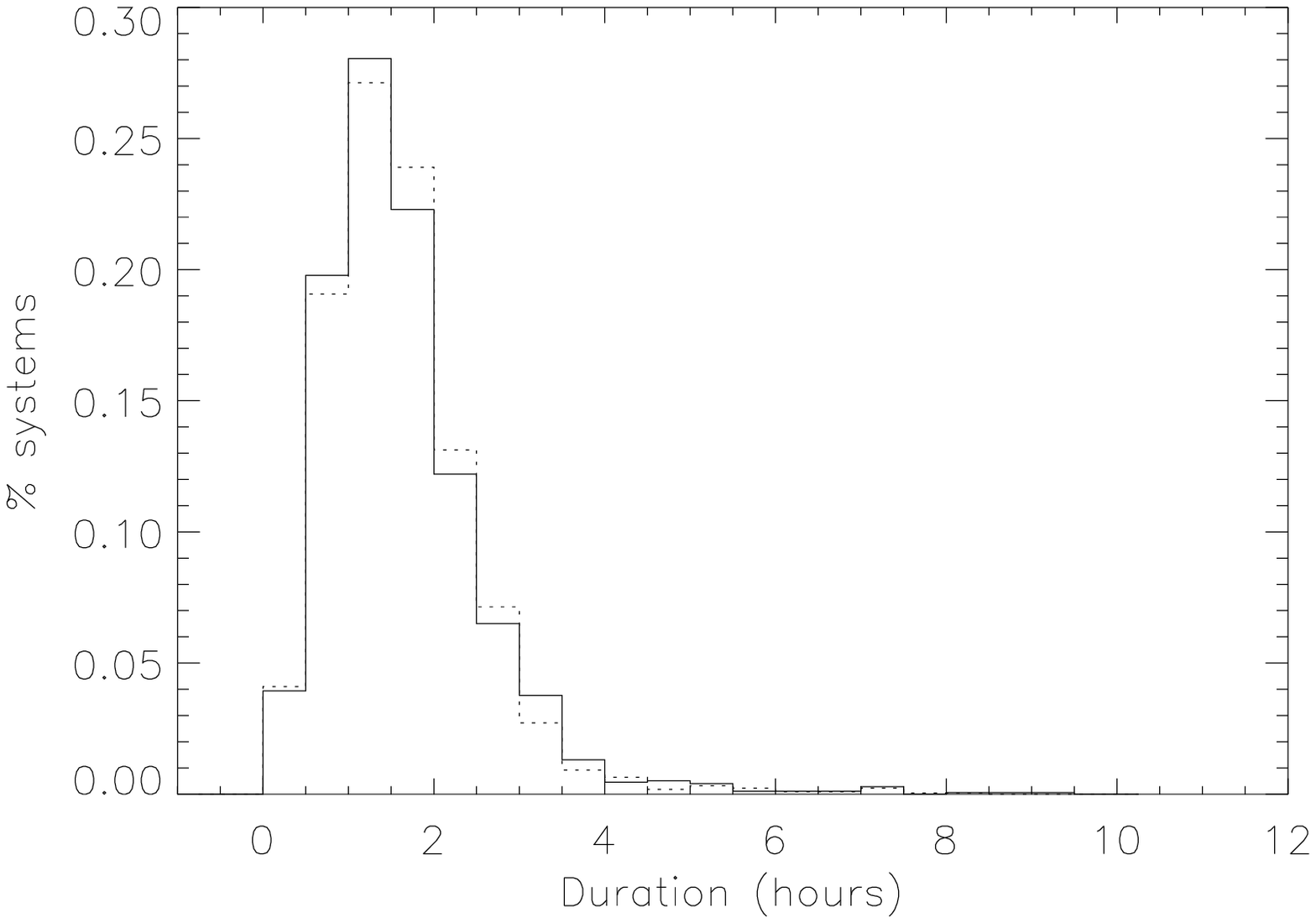,width=5truein}
\caption{The figure shows the results of our Monte Carlo simulations for the simulated eclipsing binary M-dwarf systems.  
The solid line histogram defines the results of the simulations assuming the mass-ratio distributiond defined in Halbwachs et al. (2003)
and the dotted line assumes the simulated binary systems were created by randomly paired stars from a single IMF (following DM91).
The top panel gives the distribution of maximum eclipse amplitudes, and the bottom panel shows the 
distrubtion of eclipse durations for all simulated systems detected with the observing window function. }
\label{fig:monte}
\end{figure}

We found, for both q-distributions, that most detected eclipsing binary systems in the simulation favor short period orbits,
with $\sim 85\%$ having periods $P < 10$ days.  These are on circular orbits, according to DM91.  
$75\%$ of detected systems have duration, $\tau$, between $ 1 \le\tau\le 6 $ hours.
There is a slight difference in the resulting eclipse amplitude values for the two different input mass-ratio distributions.
$70\%$ of detected systems have a maximum eclipse depth greater than $ 0.05~{\rm mag}$ when the HMUA03 q-distribution was adopted 
and approximately $60\%$ when binary systems were formed from randomly paired stars (following DM91).
We show in Figure~\ref{fig:monte} the duration and amplitude
distributions for the detected low-mass eclipsing systems in the
simulations.  Again the solid line  is the result of the simulations assuming the  HMUA03 q-distribution and the dotted 
line is the same for the DM91 q-distribution.
These results are well-matched to the characteristics of
the known eclipsing binary M-dwarfs and to the results of the simple scaling arguments above. 

\begin{figure}
\psfig{file=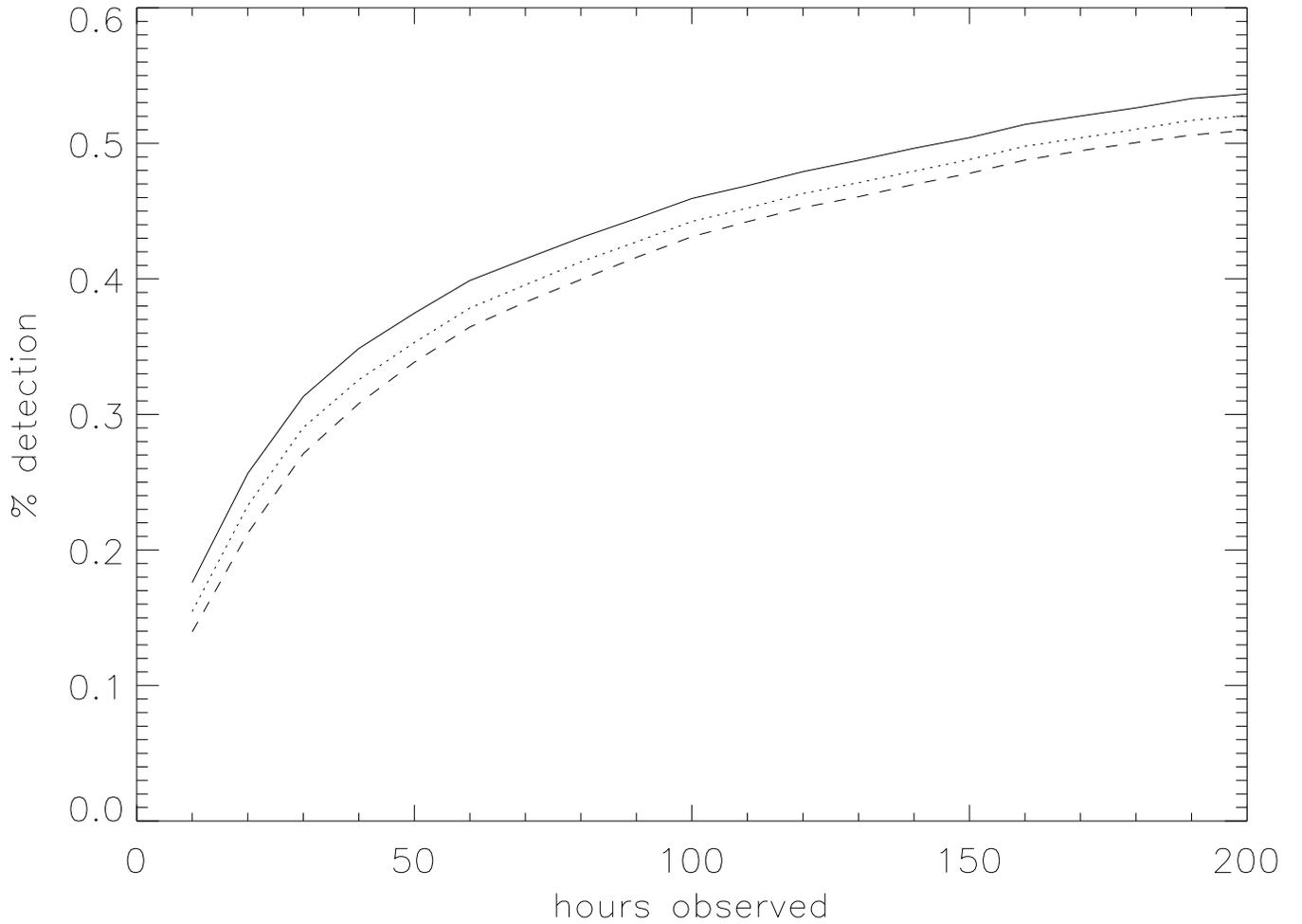,width=7truein,angle=90}
\caption{The figure shows a plot of the detection efficiency of the low-mass eclipsing binary systems defined in the Monte Carlo simulation
as a function of the simulated observing hours.  We assumed observations were taken in ten hour blocks
with a sampling rate of one per hour (dashed line), per 40 minutes (dotted line)
and per 20 minutes (solid line) for the total number of simulated observing hours on a given cluster.
}
\label{fig:hoursobs}
\end{figure}

We then tested different sampling rates and total observing hours to asses the effect of the observing
window function on the eclipse detection efficiency.  Figure~\ref{fig:hoursobs} shows a plot of the percent of the simulated eclipsing
binaries detected as a function of observing hours.  
The three different lines represent 20 minute (solid line), 40 minute (dotted) and 1 hour (dashed) sampling cadences.   
The detection efficiency was largely unaffected by whether or not the simulated observing nights were consecutive.

We thus designed our survey to achieve a photometric 
sampling rate of at least 1 pair of photometric measurements per hour (i.e. double correlated sampling).  A minimum sensitivity 
to enable the detection of a $0.05$ mag amplitude eclipse in an
average target primary star of $m\sim 0.3 M_{\sun}$ is required, although much better photometric precision is desirable
and achieved ($\sim$3 mmag precision at $I\approx 16$).  Furthermore to be consistent with evidence for the brown dwarf desert 
(Marcy \& Butler 2000; Halbwachs et~al. 2000; Zucker \& Mazeh 2001), 
we did not include short period brown dwarfs in our simulations, but if such systems exist we would be able to detect 
their eclipse signature.  A $0.05 M_{\sun}$ (age $\sim 5$ Gyr) brown dwarf orbiting a $0.3 M_{\sun}$ star in an edge-on binary system would 
eclipse producing $\sim 0.09$ magnitude amplitude signature that would be detected in our survey.

Detecting an eclipsing object requires data that
cover multiple events, which may be both primary and/or secondary eclipses; thus
we monitored the clusters on timescales of hours, days, and
months.  This provides phase and period information necessary for
follow-up observations such as radial velocity curves and more
detailed photometric coverage.  
Further, M-dwarf stars show chromospheric activity, including H$\alpha$ emission line flares,
and have sunspots which may mimic shallow eclipses, so color information,
here V and I, are optimal for distinguishing such events
from genuine eclipses. 

Finally, our simulations show that 0.08\% of the cluster stars between $0.7 M_{\sun}$ 
 and $0.2 M_{\sun}$ are eclipsing M-dwarf binaries, $\sim 70$ \% of which have amplitudes greater than 0.05
magnitudes, and a further 1/3 of those 
are detected during the simple 40 hour observing window function we adopted for the simulations.
Therefore, $\sim 1 \times 10^6$~cluster star hours are required to detect 5 eclipsing binary M-dwarfs in the open clusters.  
The total star hours required scales directly with the assumed binary fraction for M-dwarf stars 
which we have reasonably adopted to be 35\%, consistent with the value determined for field M-dwarf stars.

\subsection{Cluster sample}

The selected  open clusters  satisfy the following
criterion:  they are old enough for M-dwarfs to have settled onto the main sequence
(setting a lower limit $\ge 10^{8}$ years, Baraffe et~al. (1998), their Figure 3);
their distance modulus is small enough that M-dwarfs
 are accessible for both photometric monitoring and spectroscopic follow-up in
a reasonable time, $(M-m)_V \le 10$; the reddening is low ($E(B-V) < 0.4$); the clusters are rich enough
to allow efficient monitoring of many stars; and they are distributed in RA and DEC to allow
efficient observing during normal TAC allocations.  
Our sample of clusters covers a range of ages and metallicities, 
known from  detailed studies of the brighter members.  
The selected clusters, listed in Table~\ref{tab:cluslis}, 
have ages ranging from 200 Myr to at least 4 Gyr, and metallicity from
around solar to -0.2 dex, with apparent V-band distance moduli $8 \le (m-M)_{V} \le 10$.

\subsection{Observations}

Our survey was conducted between February 2002 and June 2003 with the Mosaic camera on the Mayall 4m telescope at
Kitt Peak National Observatory (KPNO 4m) and the Wide Field Camera on the 2.5m
Isaac Newton Telescope (INT WFC) La Palma, Canary Islands.  The KPNO Mosiac camera 
consists of eight 2048$\times$4096 pixel EEV CCDs arranged in a $4\times 2$ grid with 0.5-0.7mm spacing
between adjacent chips.  The pixel scale is $\sim 0.26^{\prime\prime}/{\rm pix}$ in the center and
decreases by $\sim$6\% out to the edges of the array, thus the entire array covers
a $36^\prime \times 36^\prime$ FOV.
The INT WFC is also a mosaic camera, consisting of four $2048 \times 4096$
pixel EEV CCDs, arranged in an L-shape, as a square 6K$\times$6K pixel pattern with a 2K$\times$2K corner cut out.  
The pixel scale is $0.33^{\prime\prime}/{\rm pix}$ with $\sim$ 1\% deviation to the edges of the detector.  
The array covers an $\sim 34^\prime \times 34^\prime$ FOV.  

\begin{figure}
\psfig{file=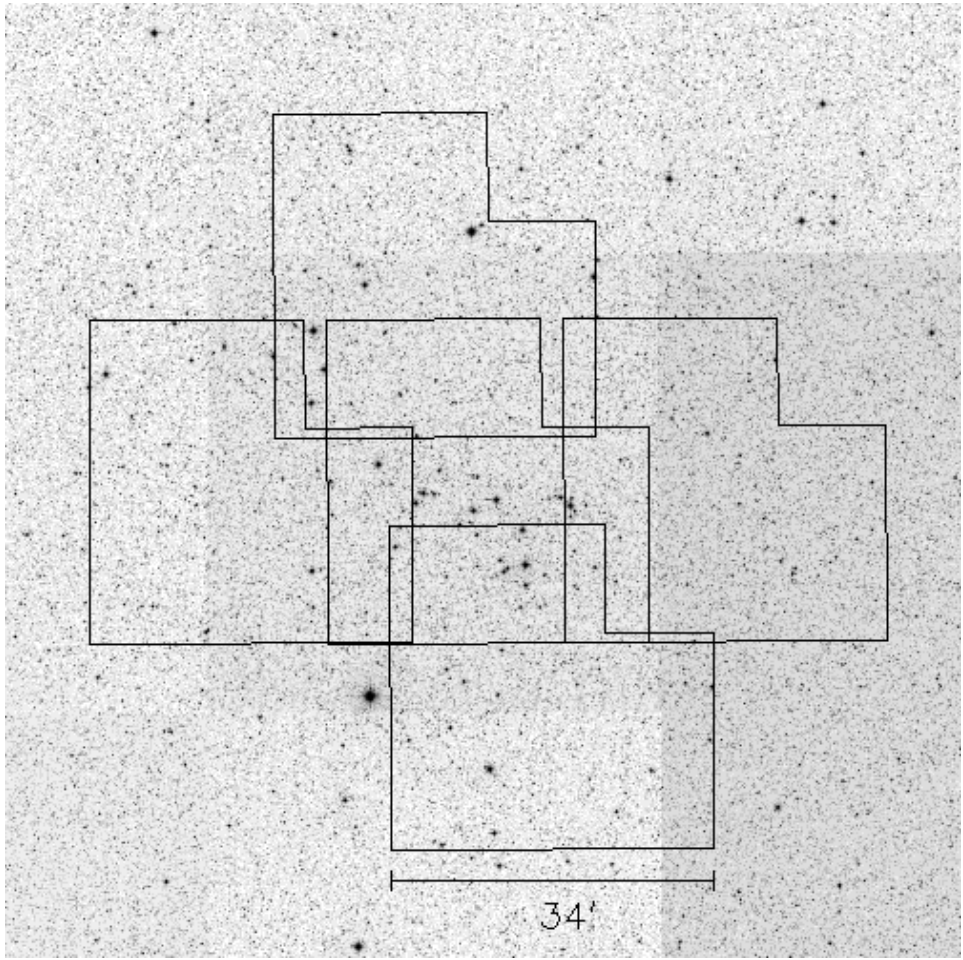}
\caption{Digitized Sky Survey in the field of NGC 6633 with INT WFC pointings. }
\label{fig:n6633}
\end{figure}

The typical core diameter of our sample clusters, measured from the
brighter stars, is $\sim 0.5^{\circ}$, which is well matched to the
FOV of the detectors described above.  We designed our survey to
sample both the inner and outer regions of the clusters where, if
mass segregation is effective, we may find significant numbers of
low-mass stars.  Our observing pattern consisted of one pointing
on the center of the cluster and four flanking fields which overlap
the central pointing by $\sim 25 \%$; this overlap provided a higher
sampling rate for a significant subset of the central region.  In
this way, we were able to monitor a $1.2^{\circ} \times 1.2^{\circ}$ FOV
with the KPNO Mosaic and a $1.0^{\circ} \times 1.0^{\circ}$ FOV with the
INT WFC.  Figure~\ref{fig:n6633} shows an image of the field of cluster
NGC~6633 with the INT WFC footprint outlining the target area.  All the
clusters were observed in both Johnson V and either Cousins I or Gunn i.

Further details of the observations depend on the specific cluster
and on the filter and instrument choice.  This information 
(including cluster, observing dates, instrument, exposures times, filters, total frames taken, and average sampling rate)
is summarized in Table~\ref{tab:obs}.

The detailed observing plan is described below.  
The five overlapping pointings on each cluster were observed in sequence throughout
the night until the cluster fell below an airmass of two.  The sequence differed
at the INT and KPNO due to the long readout time of the KPNO-Mosaic detector (2.6 minutes) 
and the need to maintain the 1 $\rm{hour}^{-1}$ mean sampling rate.
At the INT, a pair of exposures was taken at each pointing during a complete covering of the cluster region.
At KPNO, we obtained a pair of exposures on the central cluster 
field two times and on each flanking field once in the sequence that
was repeated over the course of the night.  We took pairs of exposures at each pointing in order to
provide independent confirmation of each photometric point.
Furthermore, we did not dither the observations; thus each object 
falls on the same pixels for all observations of a given cluster pointing.

At each field center pointing, we took single short exposures to provide high S/N data for 
solar-type stars without saturating them,  and pairs of longer exposures to 
provide the requisite data for 
 the fainter K and M-dwarfs which are our primary targets.  We obtained at least one complete set of 
exposures at each field center in each of the two filters used, in order to
make a complete color-magnitude diagram of all the clusters. Subsequent repeat observations of
each central cluster pointing were taken in both V and I (or i), but the flanking fields were
repeated only in the I (or i) filter.

The weather was generally good during our observing runs.  We observed
23/35 photometric nights and lost only 7 full nights due to very
poor weather.  Only differential photometry is necessary for the
primary science goals of this survey, but standard star fields were
observed on photometric nights during each of the runs to provide
calibrated CMDs.  It should be noted that standard star
sequences are available in some of our clusters, facilitating
this calibration. 

\section{Data Reduction}
\label{sec:dr}

\subsection{INT WFC Data}

The INT data were processed through the standard INT Wide Field Survey Pipeline 
created by the Cambridge Astronomical Survey Unit (CASU) and designed specifically as
a data reduction tool for INT-WFC mosaic images.
The pipeline processing treats each CCD in the mosaic independently and includes
bias subtraction, correction for the non-linearity in the CCDs (see http://www.ast.cam.ac.uk/$\sim$wfcsur/foibles.php), 
bad pixel interpolation, flat fielding, defringing, and gain 
correction which scales the whole mosaic to a uniform background level.
The same processing steps were applied to the data collected during each observing run.

For each run, the first step was to combine a large number,
typically $\sim 25$, of individual bias frames into a single 2-D
master bias image for the entire run.  All science and calibration
frames were debiased using this stacked master bias image 
together with  the overscan section associated with each chip.  Any
known bad pixels were then flagged and interpolated over.  The frames
were then trimmed and corrected for the non-linearity in the CCDs.
Master twilight flat-field images for each entire run  were
created in  each of  V and (I or i) 
using the following procedure: fifteen individual images
for each filter were averaged using a $5\sigma$ clipping algorithm,
and the appropriate stacked flat-field was then scaled and divided
into each data frame.  After applying the flat-field, a fringe pattern
was visible in those I-band images with exposure times longer than
7s (shorter exposures have a background dominated by random
noise).  In general, our science
observing program did not provide enough diverse frames
with which to create a master fringe frame with sufficient
signal-to-noise since we did not dither the individual observations.  Thus, for all runs
after the first (July 2002) we observed non-cluster dark sky fields
each night which were combined into a master fringe frame for that
run.  We averaged $\sim 10-12$ images of distinct dark parts of the
sky using a $3\sigma$ clipping algorithm to remove objects.  The only
exception was the July 2002 observing run during which no dark sky
images were observed and for which we used a fringe frame taken from
the INT WFC archives (dated February 2002).  The appropriate
fringe-map was then scaled and subtracted from all I-band program
frames for that run with exposure times longer than 7s.  Finally, all
four CCDs in the mosaic were scaled to the same mean background sky level.

At the completion of the processing, we examined many individual
images by eye and observed no residual flat-field or fringe features.    
In order to obtain a more quantitative measure
of the degree of residual flat-field and fringe 
variation left in the images after the data reduction we tested
a sample of 10 images.  We generously masked out all objects detected 
at $1.5\sigma$ above the background until we obtained a sample of pixels containing
only sky.  We compared the standard deviation of the sky values for the entire chip with the expected sigma assuming
only poisson counting errors.  On all 10 test images, we determined the sky variation to be $\le 1\%$.  

\subsection{KPNO-4m Mosaic Data}

The Mosaic data were reduced using both the IRAF\footnotemark[1]/MSCRED (Valdes 1998) package and a toolkit 
based on the INT-WFC pipeline that has been adapted by CASU for use in processing all types 
 of multi-extension fits files generated by mosaic detectors.  The same data processing
steps were used for each observing run and are outlined below.

\footnotetext[1]{IRAF is distributed by the National Optical Astronomy Observatory, which is operated by the Association of Universities for Research in Astronomy, Inc., under
cooperative agreement with the National Science Foundation.}
    
The first correction made to the raw Mosaic data was to remove the ghost signatures of
saturated stars in four of the eight CCDs due to crosstalk in the controller electronics.  
Crosstalk files provided by KPNO are available for this purpose.  We used calibration files
generated from data taken on August 30, 2002 to correct our September 2002
and January 2003 data, and files generated on March 28, 2003 were used on our February
2003 and June 2003 data.  Next, for each run we combined 20-25 individual bias frames into a single 
master 2-D bias image.  The master bias frame was subtracted from all 
Mosaic images which were then overscan corrected and trimmed. 
We combined $\sim 15$ twilight flats taken over the course of each run into a master 
twilight flat for each filter, which we then applied to the science frames.  

For observing runs after September 2002, we took specific non-cluster images to use in creating dark-sky flats.  
We observed $\sim 15$ dithered dark-sky images and processed them with the program frames.
The IRAF {\it mscred.objmasks} task was used to create object masks for each individual dark-sky
frame.  The masks were used when combining the individual images into a master dark-sky flat in each filter with the 
{\it mscred.sflatcombine} task.  Finally, we smoothed the master dark-sky flat with a 7x7 pixel box filter. 
The 7-pixel box was small enough so as not to smooth out any residual flat-field features, but large enough to 
reduce the noise in the final dark-sky flat to the level of a typical flat-field.  
We did not apply dark-sky flats to any exposures $ < 5 {\rm s}$ as the background level was very low and the application
of the dark-sky flats only added noise.

The Mosaic images taken at the KPNO-4m had an instrumental feature in
the I-band images, a scattered light image of the pupil. The
scattered light image is an additive feature, not a multiplicative
feature, so cannot be removed with the application of the flat-field.
We tested the MSCRED tasks {\it mscpupil} and {\it rmpupil}, and the
CASU toolkit routine {\it fitsio\_scatter}, designed to remove the
feature.  The CASU toolkit routine was consistently better able to fit and remove 
the low level feature in our crowded star fields.
The routine performs a post-flat-field correction to the
data image, $D$, by performing the following arithmetic on the image:
$D^{\prime} = D * P - (P-1)*medc(D)$ where $P$ is the normalized pupil
image, $medc(D)$ is a $3\sigma$ clipped measure of the median sky
value of the data image, and $D^{\prime}$ is the corrected output.
The input pupil image was created by simply dividing a smoothed
dark-sky flat by the same dark-sky flat which was removed of its
scattered light with the {\it mscpupil} task and then smoothed.

At this stage of data reduction, fringing was still visible in the I-band images.
The same science frames used to create the
dark-sky flat -- now post both twilight and dark-sky flat application
-- were combined with object masks into a final fringe frame which was
then scaled and subtracted from all the I-band science images with
exposure times $> 5 {\rm s}$.  After completing all the data
processing, we tested a sample of 10 Mosaic images in the same fashion
as described above for the INT data and found residual
instrumental signatures left in the Mosaic data to be $\le 1\%$.  

Any residual instrumental variations should not affect our ability to detect eclipse signatures.
Although the large scale residual features are of amplitude 1\%, 
we are performing differential photometry locally on each CCD, thus only systematic spatial variations
on scales on the order the offset between individual exposures (typically less than 10 pixels) are relevant.  These
are far less than any large scale residual features as seen by our final photometric precision discussed below.  

\subsection{CASU catalogue extraction software: Source Detection, Astrometry and Classification}
\label{ssec:casu}

At this point, all instrumental signatures have been removed from the
data.  All science frames taken at the INT and KPNO were then
processed using the CASU catalogue extraction software which
includes source detection, aperture photometry, astrometry and
morphological classification
(http://www.ast.cam.ac.uk/$\sim$wfcam/documentation.html; Irwin \& Lewis 2001).

The source detection algorithm defines an object as a set of contiguous pixels above a defined detection threshold.  
The routine requires as input the detection
limit in units of background $\sigma$ and the minimum number of connected pixels above that threshold which define an object.
Since we have multiple exposures of each field with which to establish the reality of a detection, we can set 
a low detection threshold of $1.5\sigma$ and a small minimum source size of 5 pixels, while requiring that 
all `real' objects be detected in both the V and I (or i) band stacked master images.
We applied confidence maps, which define dead pixels and other problems on the detector, with the detection  
routine in order to limit further the number of false detections.   
We also employed the deblending option which attempts to deconvolve any
overlapping objects.  For each detected
object, aperture photometry is performed with five discrete apertures,
and these measurements are used in a curve-of-growth analysis to
define the total flux in the object (Irwin 2001; Hall \& Mackay 1984).
We chose the value of the core aperture to be 4~pixels, 
which corresponds to $1^{\prime\prime}$ on the KPNO images
and to $1.3^{\prime\prime}$ on the INT images.  
All flux  measurements are corrected for the varying pixel scale across the detector, which 
is significant for the wide-field detectors we are using ($\sim 1\%$ for INT-WFC and $\sim 6\%$ for the KPNO-Mosaic).     
Objects are then classified as stellar, non-stellar, or noise, based on the combination of radial profile and ellipticity.  

We will derive light curves for each detected object based on an
automated identification technique that finds the same object in
subsequent frames based on its coordinates in the world coordinate
system (WCS). For this, the initial astrometric solution present in
the image headers is not sufficient. 
We refined it iteratively by comparing the positions of detected objects with those in the USNO-A V2.0 catalogue of
astrometric standards (Monet 1998).   We obtained  an
updated astrometric solution with an internal precision of $\sim 100
{\rm mas}$ and an external precision limited by the USNO data of $\sim
0.25^{\prime\prime}$. 

\subsection{Calibration}

Although only differential photometry is required for our primary
science goals, we did observe several standard star fields during each
photometric observing night including the standard star fields in two of our target clusters, M67 and NGC~6940.
The standard fields were processed along with the science
frames as described above, and aperture photometry, with a 4 pixel aperture, was performed on
the processed images.  An aperture correction
derived from the curve of growth analysis was added to the
instrumental magnitudes.
We transformed the observed standard stars according to the following equations:  
$$ V = a_0  + v_{inst} + a_1*(X-1) + a_2*(V-I)$$ $$I = b_0
+ i_{inst} + b_1*(X-1) + b_2*(V-I)$$ where $V,I$ are the catalogued magnitudes 
of the standard stars, $v,i_{inst}$ are the instrumental magnitudes, and
$X$ is the airmass.  

For the KPNO Mosaic images we used the {\it digiphot.photcal.fitparams} task in IRAF to solve for
the coefficients, $a_i$ and $b_i$.   
In September 2002 and June 2003, we observed Stetson (Stetson 2000)
photometric standard fields which extend the Landolt (Landolt 1992)
standards to fainter stars with a larger color range ($1.0 \lesssim
V-I \lesssim 3.0$).  In January and February 2003, we observed the
standard star sequence in the M67 field defined by Montgomery et~al. (1993).
We used only the Stetson standards, which have a large color range, to solve
for the color terms, $a_2$ and $b_2$.
We observed 5-25 standards on each chip from at least one of the fields
L92, L95, L110, L113 or the standard field centered on NGC 6940 and
determined the appropriate color terms for each individual CCD chip.  The color
terms derived independently for the two observing runs are shown in
Table~\ref{tab:cterms}, along with their mean value.  We applied these
average values to the transformation equations and solved for the
atmospheric extinction terms and zero points for each photometric night (given in
Table~\ref{tab:extterms}).

\begin{figure}
\psfig{file=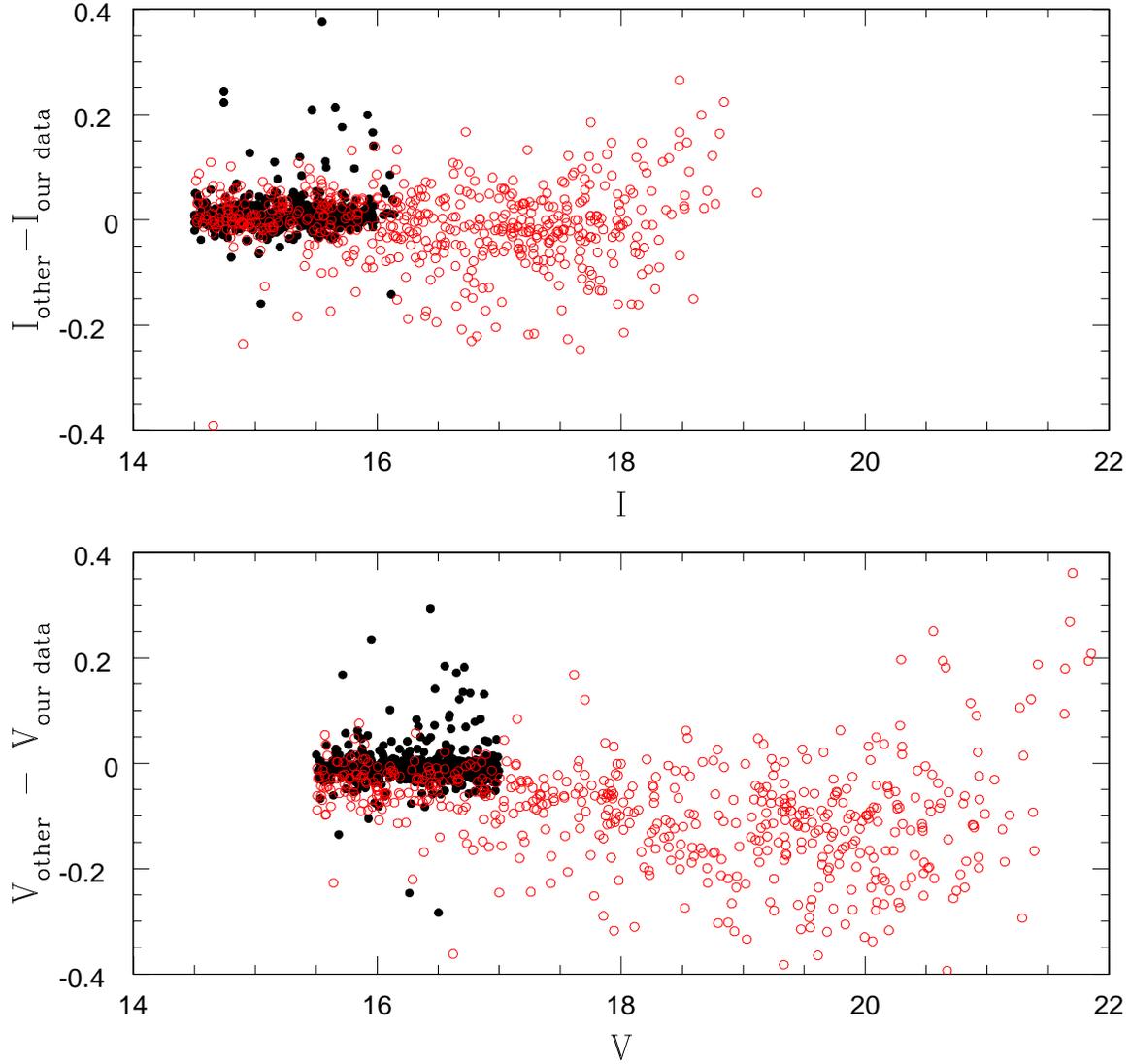,width=6truein}
\caption{Comparison of central M35 field observed January 2003 with other published datasets of the same field.  Solid cirles -- Table 1 of Sung \& Bessell 1999 compared
with our data.  The handful of poorly matched points are likely blended objects.  Open circles -- Table 2 of Barrado~y~Navascu\'es et~al. 2001 compared with our data.  The bright end cut off is due at the saturation level of our data. }
\label{fig:compphot}
\end{figure}

As a test of the photometric calibration, we cross correlated the
positions of objects detected in the central NGC~2168 (M35) field
against two previously published datasets from Sung \& Bessell 1999
(SB99) and Barrado~y~Navascu\'es et~al. 2001 (B01).
Figure~\ref{fig:compphot} is a plot of the magnitude difference
between our data and the published sets for objects matched to 
 less than 1~pixel ($0.25^{\prime\prime}$). 
The comparison with the SB99 data shows a
tight correlation over the limited overlapping magnitude range and no
significant slope as a function of either V or I magnitude.  The small
mean zero point offsets ($I_{offset} = 0.007 $ and $V_{offset} =
-0.012$) are within our calibration errors.  We used the median absolute deviation (MAD) about the median as
a robust estimate of the standard deviation (for a normal distribution $\sigma \sim 1.48\times $MAD).   
The standard deviation estimate in each band is small ($\sigma_I = 0.031$ and $\sigma_V = 0.044$).
We also compared our data to the observations taken at KPNO
and published in B01.  These data show
an insignificant mean offset from our data of $I_{offset} = -0.0045$ in the I-band,
but a larger standard deviation of $\sigma_I = 0.099$.  The B01 V-band
measurements are offset from our data by $V_{offset} = -0.076$
and have a standard deviation of $\sigma_V = 0.129$.
Our calibrations are well matched to the two previous studies for all except the V-band data of B01, but as noted in their
paper, the B01 V-band measurements are also offset from the SB99 data by $\sim -0.03$ mag.  
Thus our calibrated photometry is well-matched to all other robust measurements used for comparison.

For the INT data we used the standard extinction coefficients and color 
terms given on  the INT-WFC instrument web page, namely 
$a_1=-0.12$, $b_1=-0.04$, $a_2=+0.005$, and $b_2=+0.009$.  These are typical values for
this site. 
We then independently solved for the zero points, $a_0$ and $b_0$ for each photometric night.
 
\section{Photometry and Lightcurve Generation}
\label{sec:phot}

Candidate low mass eclipsing binary stars are detected through
variations in their light curves.  We developed a differential
photometry routine that measures variations relative to the vast
majority of stars in the field, which are assumed to be intrinsically
non-variable.  We generate light curves for all stars in the observed
cluster fields using a coordinate-based aperture photometry routine.

\subsection{Differential Photometry}

A master list of detected objects and their RA and DEC positions is
generated for each pointing and filter combination.  This list is obtained by 
first stacking five observations of a given pointing with the best
seeing, and then applying the CASU catalogue extraction routines to this
stacked image.  A master catalogue containing all detected objects and
their positions is output by the software.  We then transform the coordinates of 
objects detected on the master image into the reference frame of each 
individual image according to the transformation coefficients defined in the headers.
The coordinate transformation matrix is of the form: $$x^{\prime}_i=a*x_i+b*y_i +c$$
$$y^{\prime}_i=d*y_i+e*x_i+f.$$ where ($a$,$b$,$c$,$d$,$e$,$f$) are the transformation coefficients, 
($x_i$,$y_i$) are the coordinates for a single object detected on the master image,
and ($x^\prime_i$,$y^\prime_i$) are the corresponding coordinates for that same object in an individual frame.
All positions are corrected for the radial
geometric distortion prior to the transformation, and the final error
in the transformed coordinates is $< 0.2$ pixels.  We then perform
aperture photometry, as described in \S~\ref{ssec:casu}, on each image
at the positions defined in the master list.

The `light curves' that would be obtained from the sequence of
measurements obtained in this way contain both intrinsic variability
and variations due to other effects such as changes in the seeing, the
airmass, the zero point, and the overall transparency of a particular
image.  We need to remove all non-intrinsic effects; our underlying
assumption in deriving differential light curves is that most stars in
a given image have no intrinsic variability.  

We use the brightest stars on each image to determine the differential offset for all stars on that frame.  Further, in order to generate light curves with the smallest possible scatter we use different aperture sizes for stars of different magnitudes.
For each of the five
pointings per cluster field, we calculate the mean magnitude (with a $3\sigma$ rejection) over 
all observations of each bright isolated star ($16\lesssim V \lesssim 18$) within three
different apertures ($r_{\rm core}$, $\sqrt2/2*r_{\rm core}$, and
$2*r_{\rm core}$ with $r_{\rm core}=4$ pixels).
We then determine the $3\sigma$-clipped mean offset between the individual bright star 
magnitudes and their corresponding mean values for each CCD chip of each individual observation.
For each star in the master list, we then create a set of three (one
for each aperture) differential lightcurves by correcting the
magnitudes measured on each image with the corresponding offsets
derived from the bright stars.  We also determine the standard
deviation in each individual lightcurve.  Ideally, the three
lightcurves for each star should be the same, but in reality they
exhibit different scatter about the mean.  Thus, we bin the objects
into 0.25 magnitude bins and calculate the average standard deviation
in the lightcurves for the stars in each bin for all three different
aperture choices.  We choose the aperture value that produces the
smallest average standard deviation for all stars in that bin.  The
magnitude measurements with the chosen aperture are used in deriving
the final differential lightcurves for all stars in the bin.  In
general, the bright stars produce lightcurves with the least scatter when
measured with the largest aperture, $2r_{\rm core}$, while the
faintest stars are better fit with the smallest aperture, $r_{\rm core}$.

Finally, the stellar positions in the master lists for all five
pointings are cross-correlated using the same linear transformations
described above.  All stars observed on more than one pointing, are
matched, and their differential lightcurves are combined.

\begin{figure}
\psfig{file=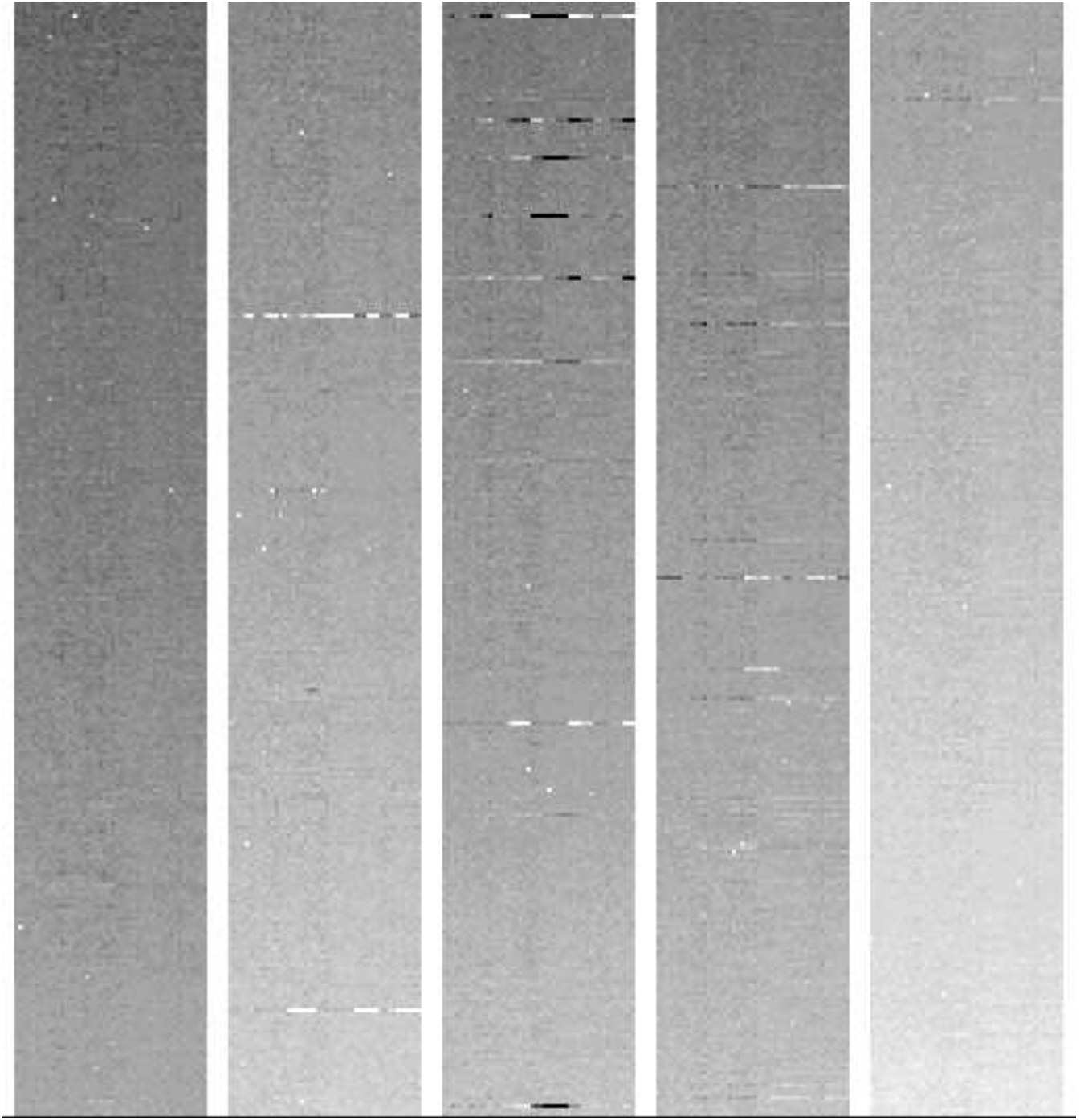,width=7truein}
\caption{A image representation of a sample of lightcurves of stars with $19 \le I \le 20$ magnitude observed in the field of the M35 cluster.  
The large columns represent 
individual CCDs on the image and each row is a differential lightcurve for a star on that CCD.  
The x-axis is observation number (not date) which are not equally spaced in time.  
Several potential variable stars are visible as with a range of amplitudes and degrees of periodicity.  See text for details. }
\label{fig:stripe}
\end{figure}

Figure~\ref{fig:stripe} shows an image representation of a sample of the differential lightcurves for stars
observed in the M35 cluster field in January 2003.
Each large column in the image represents a different CCD chip and each row contains the 
lightcurve for a single star on that chip.  
The value of a pixel corresponds to the differential magnitude of a given star on a 
single observation; the lighter pixels correspond to brighter magnitudes. 
The stars are sorted by median magnitude, therefore the overall shade of the 
column tends to get darker (fainter) towards the top of the image.  Several potential
variable stars are visible in the image as individual rows with larger than average changes in the pixel value.  
We use similar images to check for systematics in our differential lightcurve data, e.g. spurious position-dependent effects.

\begin{figure}
\psfig{file=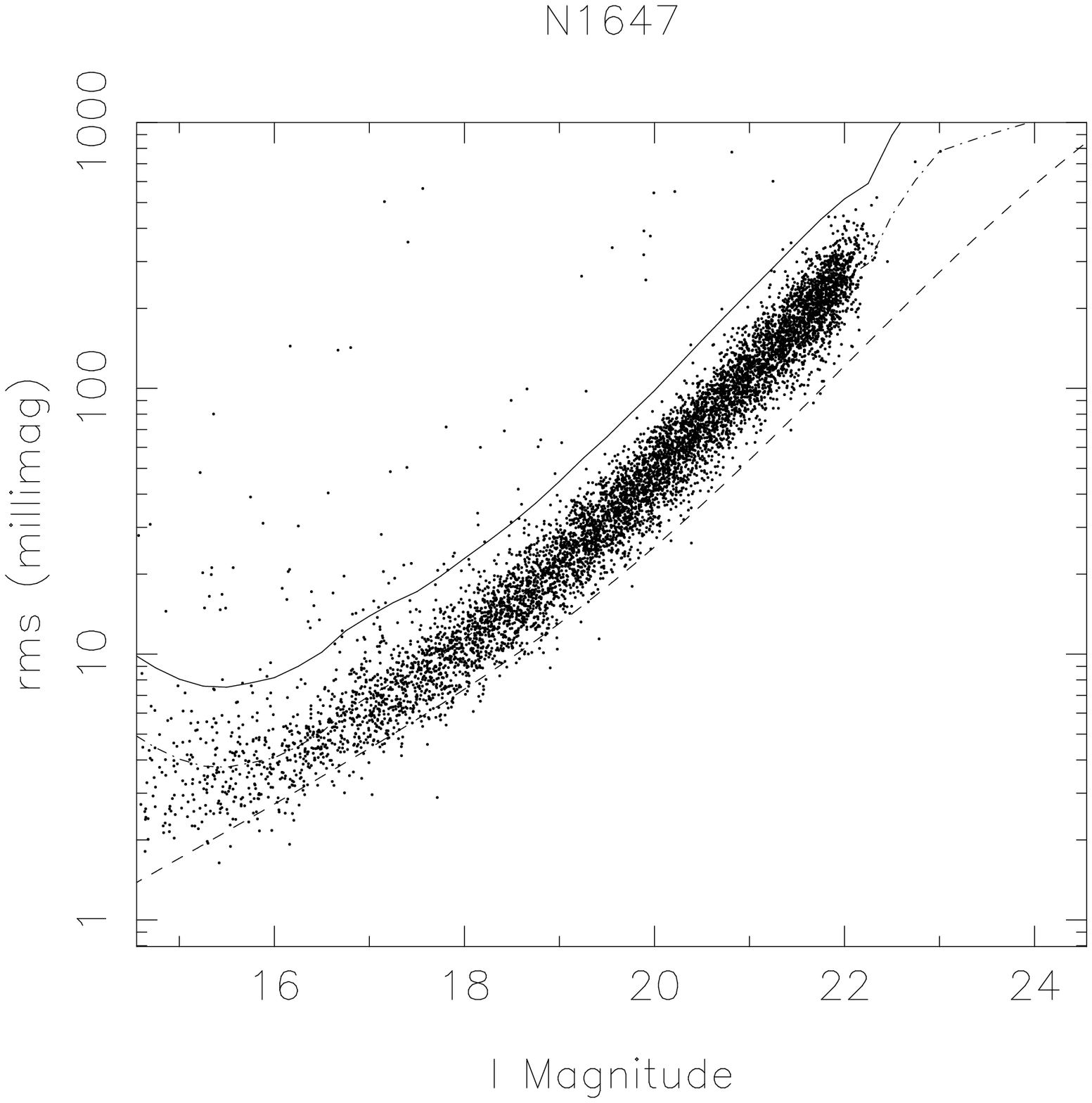,width=5truein,angle=-90}
\caption{Standard deviation in the lightcurves versus I-band magnitude for all non-blended objects in the NGC~1647 cluster field.  The dashed line
is the theoretical curve including poisson counting errors, read noise, and noise in the sky background.  The dot-dashed
line is the median of the data points as a function of magnitude and the solid line is 2$\sigma$ level above the median.
}
\label{fig:rms1647}
\end{figure}

\begin{figure}
\psfig{file=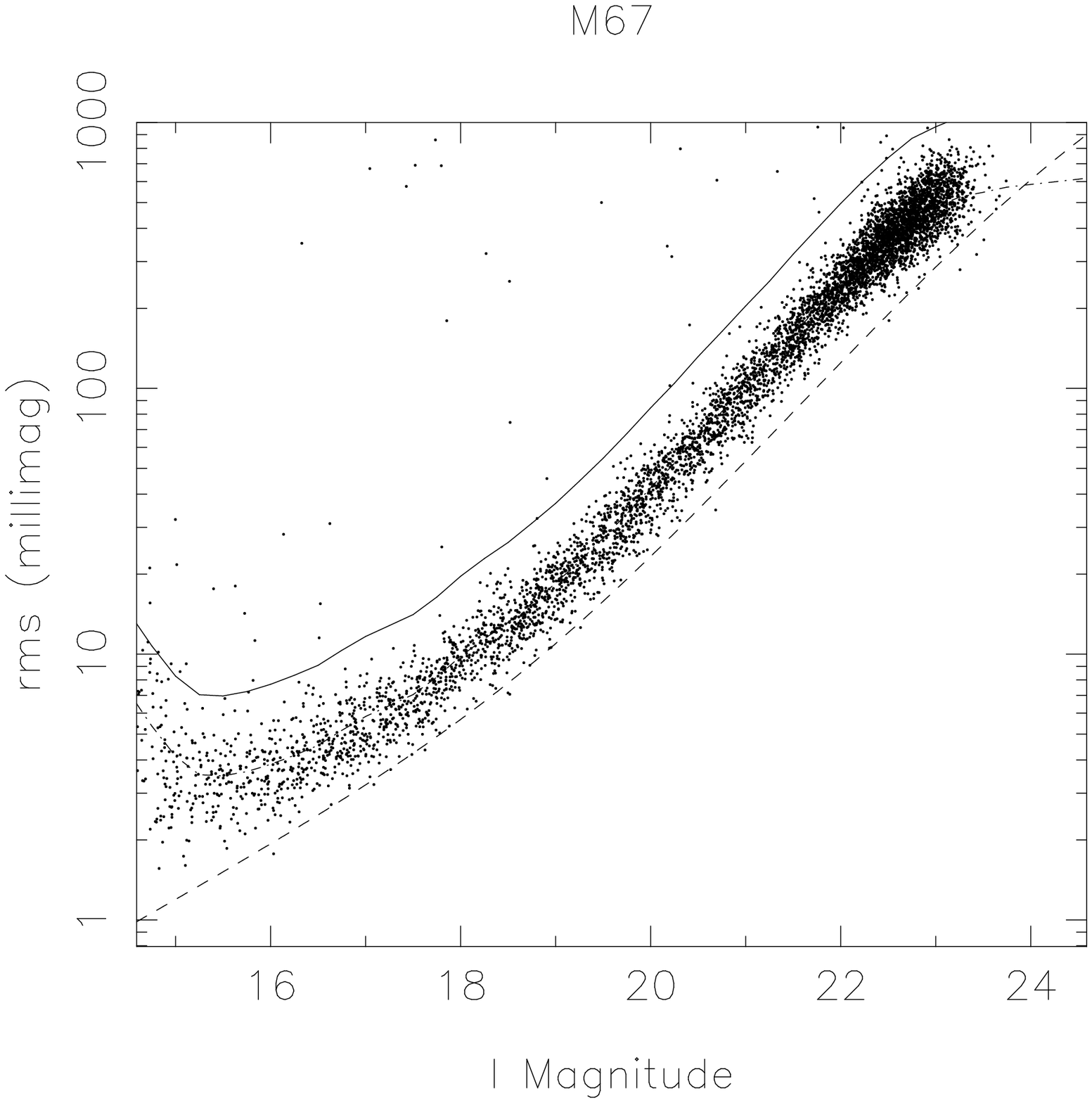,width=5truein,angle=-90}
\caption{Standard deviation in the lightcurves versus I-band magnitude for all non-blended objects in the M67 cluster field
from data observations taken at KPNO.  The dashed line
is the theoretical curve including poisson counting errors, read noise, and noise in the sky background.  The dot-dashed
line is the median of the data points as a function of magnitude and the solid line is 2$\sigma$ level above the median.  }
\label{fig:rmsm67}
\end{figure}

\begin{figure}

\clearpage

\psfig{file=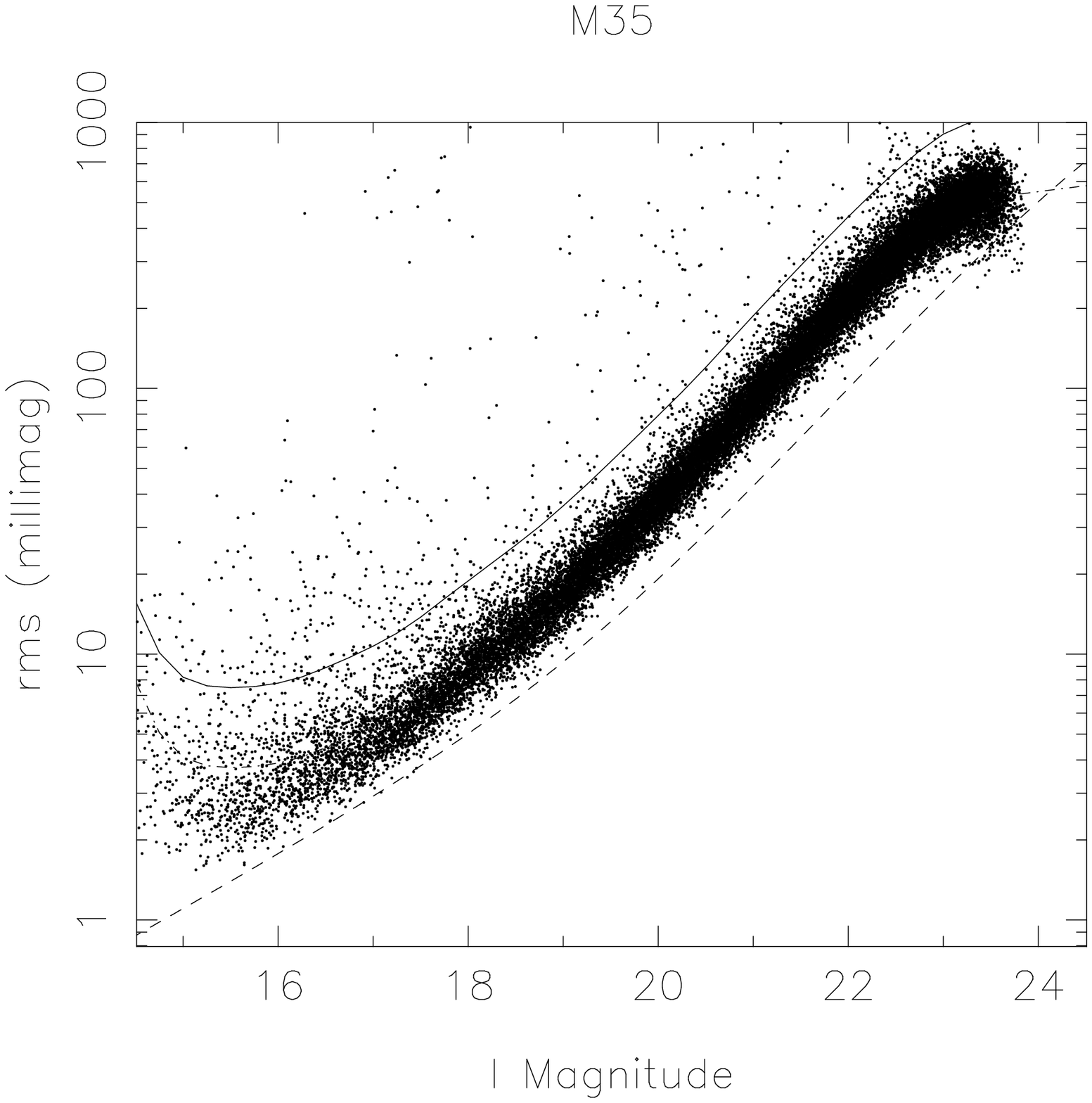,width=5truein,angle=-90}
\caption{Standard deviation in the lightcurves versus I-band magnitude for all non-blended objects in the M35 cluster field 
from observations taken at KPNO.  The dashed line
is the theoretical curve including poisson counting errors, read noise, and noise in the sky background.  The dot-dashed
line is the median of the data points as a function of magnitude and the solid line is 2$\sigma$ level above the median.  }
\label{fig:rmsm35}
\end{figure}

\begin{figure}
\psfig{file=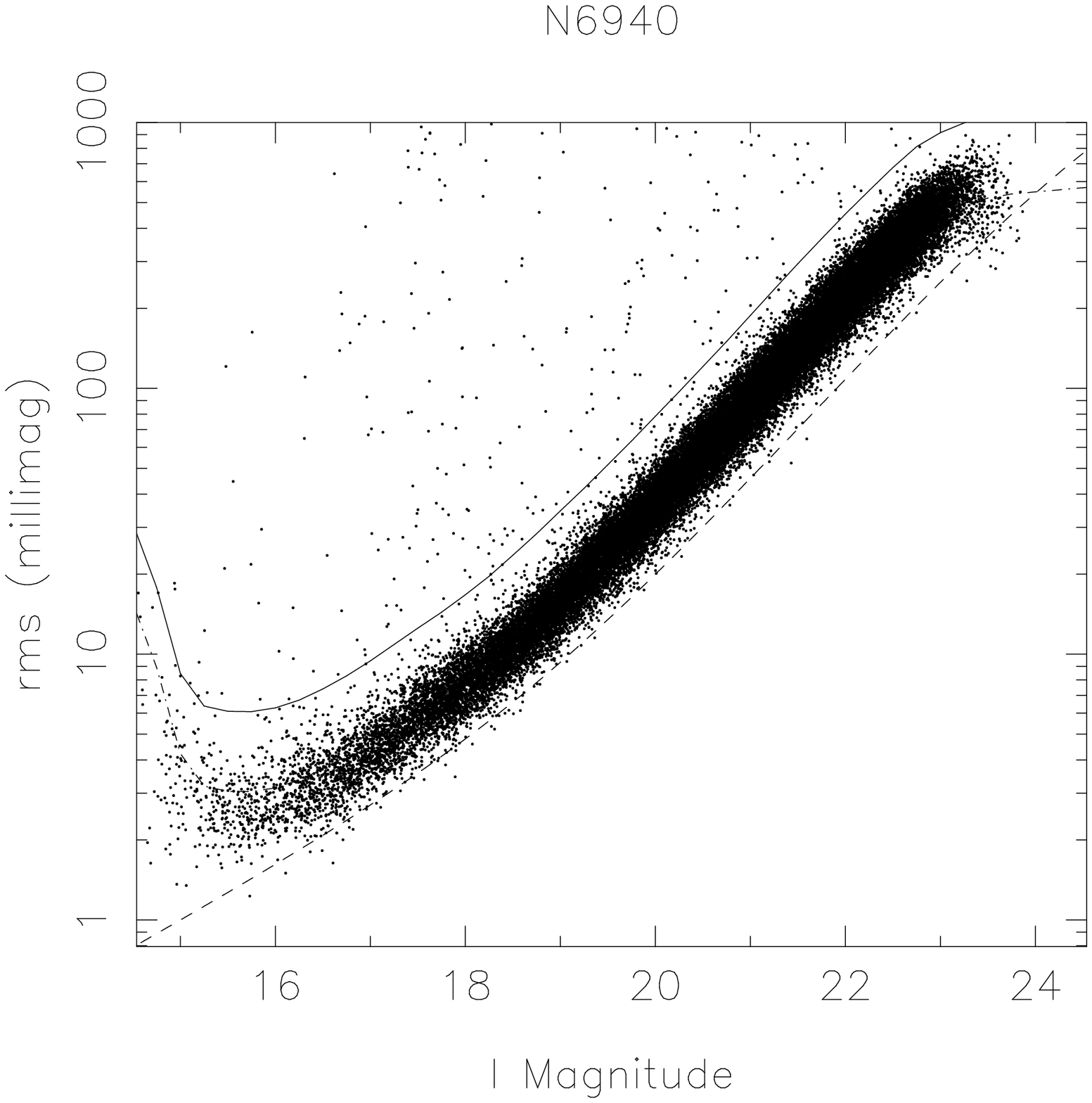,width=5truein,angle=-90}
\caption{Standard deviation in the lightcurves versus I-band magnitude for all non-blended objects in the NGC~6940 cluster field.  The dashed line
is the theoretical curve including poisson counting errors, read noise, and noise in the sky background.  The dot-dashed
line is the median of the data points as a function of magnitude and the solid line is 2$\sigma$ level above the median.  }
\label{fig:rms6940}
\end{figure}

The precision of our differential photometry may be demonstrated by plotting the standard deviation in each lightcurve 
as a function of magnitude, for all detected stellar objects (field stars and potential cluster stars).  
Figures~\ref{fig:rms1647} through \ref{fig:rms6940} 
show the standard deviation for each stellar object as a 
function of calibrated I-band magnitude for four different clusters (NGC~1647, M35, M67 and NGC~6940).
Deblended objects are flagged and excluded from the diagrams.  The dashed line is the theoretical estimate of the
scatter including poisson counting errors, read noise, and noise in the sky background.    
The dot-dashed line in the figures shows the median of the distribution as a function of magnitude.  
The difference between the theoretical curve and the observed scatter is a measure of other systematic errors in our data.
The solid line represents a $2 \sigma$ cut above the median error level; stars falling above the solid line
likely include many real large-amplitude variables, as well as some intrinsically non-variable objects which have
variable lightcurves due to defects in the detector or other sources of error in the data.

\subsection{Calibrated Photometry}

All the aperture photometry measurements observed on photometric nights were then calibrated with the 
appropriate zero point and extinction term.  
We averaged the multiple observations of each star in the master list into a single calibrated measurement. 
We defined the average error in the magnitude of each object to be the standard deviation in these measurements.
Finally, we matched the V and I master catalogues of each cluster field, based on the object positions, and 
applied the appropriate color correction
term to all objects detected in both bands.  These data comprise the final calibrated object catalogs. 

\begin{figure}
\psfig{file=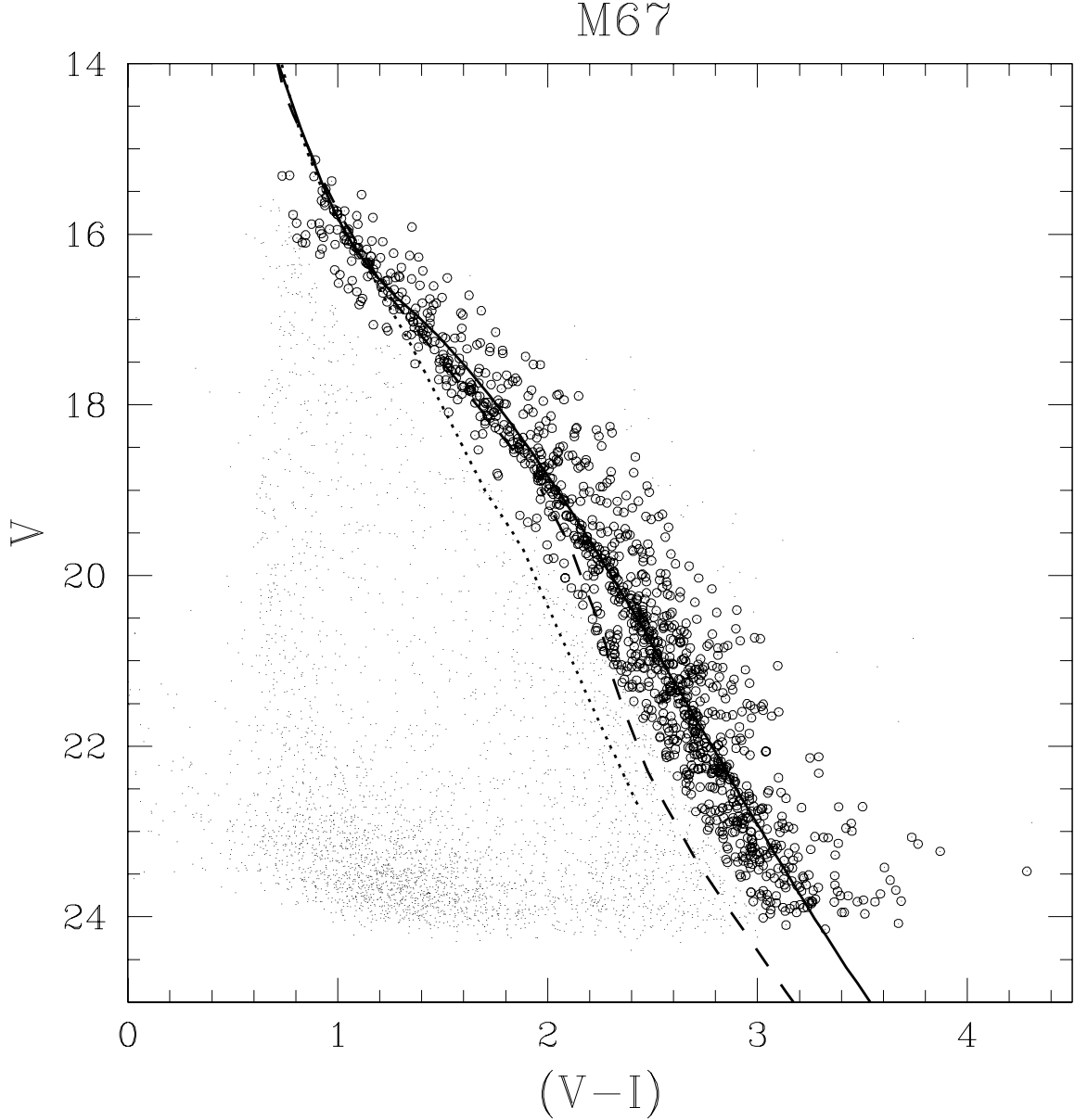,width=6truein}
\caption{Empirical Color-Magnitude Diagram for all matched objects with stellar morphological classification in the central field of M67 cluster ($36^\prime \times 36^\prime$ FOV).  The open circles mark potential cluster members chosen in a band around the
fiducial sequence defined by the
error in the calibrated magnitudes plus $0.2$ magnitudes of color to the blue and $0.5$ to the red.  The solid line is the empirical
fiducial main sequence for the solar neighborhood defined by Reid \& Gilmore (1982).  The dotted line is the 4 Gyr, solar metallicity isochrone
 from Girardi et~al. (2000) and the dashed line is the 4 Gyr, solar metallicity isochrone from Baraffe et~al. (1998) with the ${\rm L_mix}$ parameter
chosen to fit the sun (${\rm L_mix} = 1.9 {\rm H_p}$).  We applied distance and reddening estimates, $(m-M)_0=9.6\pm 0.03$ and $E(B-V)=0.04 \pm 0.01$ for the cluster from 
Sandquist (2004).}
\label{fig:m67cmd}
\end{figure}

The calibrated master catalogues may be used to determine which objects to classify as candidate cluster members.  
The $V$ versus $V-I$ color-magnitude
diagram for the central pointing of the M67 cluster is shown in Figure~\ref{fig:m67cmd}.  
We defined a fiducial cluster sequence by eye which matched the obvious feature in the diagram. 
Candidate cluster objects were chosen around the fiducial sequence in a band defined by the
error in the calibrated magnitudes plus $0.2$ magnitudes of color to the blue and $0.5$ to the red in 
order to include any objects on the binary sequence.  We also show the empirical spline fit to the sample of solar neighborhood dwarfs
defined in Reid \& Gilmore (1982) and two 4 Gyr, solar metallicity theoretical isochrones from Girardi et~al. (2000) and Baraffe et~al. (1998).
We adopted distance and reddening parameters of $(m-M)_0=9.6\pm 0.03$ and $E(B-V)=0.04\pm 0.01$ 
for the cluster from Sandquist (2004).  Although the Baraffe et~al. (1998) isochrone is a better fit to the data, neither theoretical
model is a very good match to the data at the low-mass end of the main sequence.  


\section{Variability Detection}
\label{sec:var}

\begin{figure}
\psfig{file=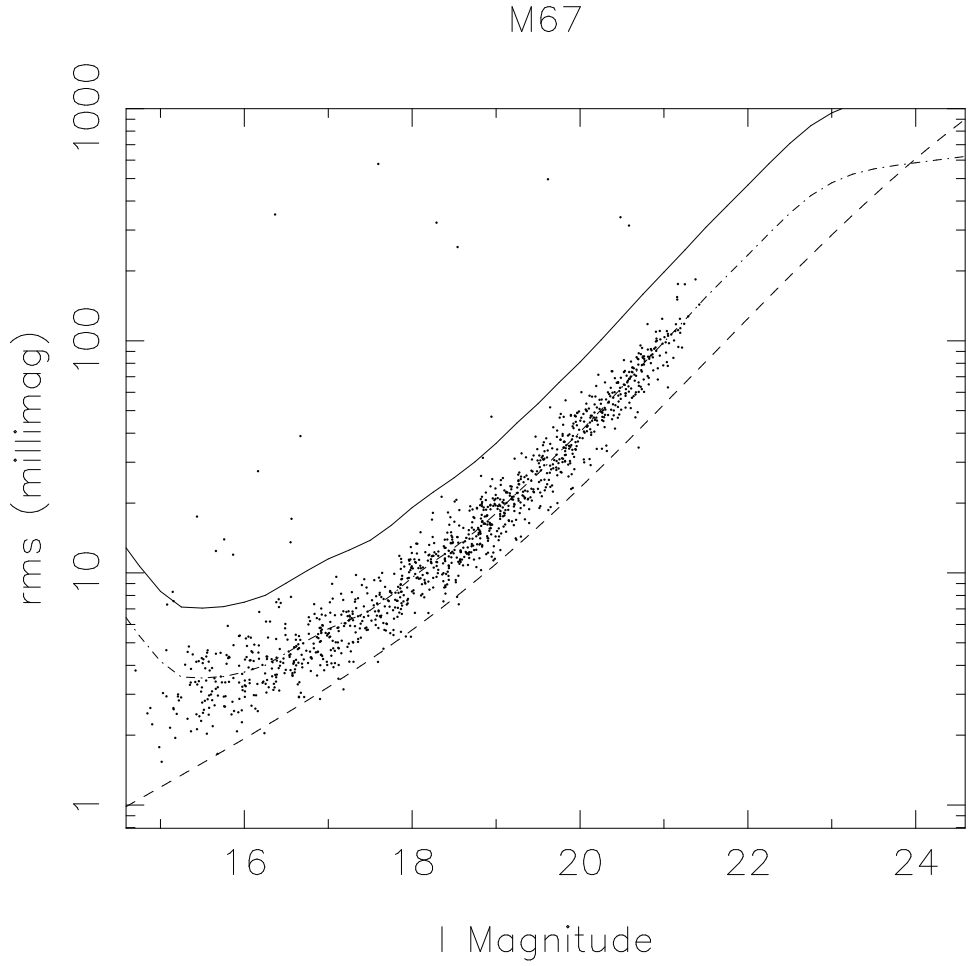,width=5truein,angle=-90}
\caption{Standard deviation in the lightcurves versus I-band magntude for all non-blended objects defined as potential cluster members.  
The lines are defined as for Figure~\ref{fig:rmsm67}.}
\label{fig:m67rmsclus}
\end{figure}

\begin{figure}
\psfig{file=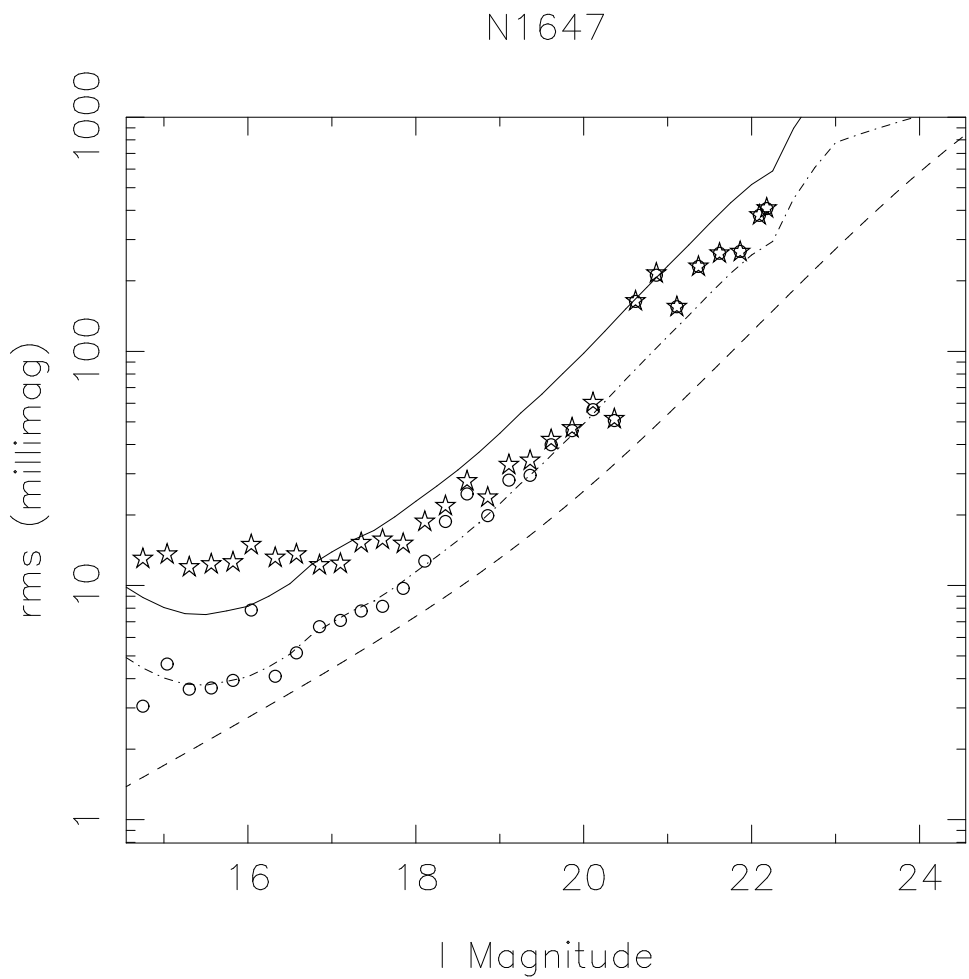,width=5truein,angle=-90}
\caption{Standard deviation vs. I magnitude for a subset of lightcurves in field of the NGC~1647 cluster.   Open circles mark the rms value for a randomly chosen subset of stars which cover the range in magnitude of the entire sample.   We added an eclipse signal with an amplitude of 0.05 magnitudes, a period of 1.6 days, and a duration of 2 hours to each lightcurve in the subsample and plotted the new rms values as large open stars.  The lines are the same as shown in Figure~\ref{fig:rms1647}:  the dashed line is the theoretical calculation of rms vs. magnitude for the original sample of NGC~1647 stars; the dot-dashed
line is the median of the data points as a function of magnitude; and the solid line is 2$\sigma$ level above the median.  
}
\label{fig:rms_addeclipse}
\end{figure}

In order to determine which objects are potential eclipsing binaries, 
we first examine the plots similar of lightcurve rms versus
magnitude for the stars in each cluster.  Figure~\ref{fig:m67rmsclus} shows such a plot for the stars defined as potential cluster members for M67 
(marked by open circles in Figure~\ref{fig:m67cmd}).  Objects with rms $> 2\sigma $ above the median error level are potential variables.  
Upon further examination, these objects include 
long-period variables and continuously varying objects both of which deviate from their median brightness over a significant
portion of their lightcurves.  Also included are objects with large-amplitude deviations from their median brightness.  These 
large variations can be due to either intrinsic stellar variability or artificial sources in the data.  Low mass eclipsing
binary systems which show short, low-amplitude variations in brightness in a majority constant lightcurve do not
exhibit large rms values associated with their lightcurves.  In order to illustrate this point, we randomly selected from our
sample a subset of objects which cover the magnitude range of the entire sample.  
We added an eclipse signal with a 0.05 magnitude amplitude, a duration of 2 hours, and
a period of 1.6 days to their lightcurves.  We plot the rms values of the original and eclipse-added lightcurves 
in Figure~\ref{fig:rms_addeclipse}.
We mark the rms value of the original lightcurves with an open circle
and the rms value of the same lightcurve with the artificial eclipse signal added as an open star.  
With the addition of the eclipse signal, the global rms value of the lightcurves change by several mmag.
The amplitude of the added eclipse is an order of magnitude greater than the standard deviation in the brighter stars, thus 
the addition of the eclipse signal has a significant affect on the rms value at the bright end.
Low-mass eclipsing binaries, on the other hand, are still indistinguishable from
typical non-variable objects in this data.  Therefore, we consider other approaches to detecting eclipse signatures in the lightcurves.  

\subsection{Stetson J-Index}
\label{sec:stetsonJ}

We designed our observing strategy so that pairs of observations were taken at each field pointing in order to provide immediate confirmation of each photometric measurement.  
This enabled us to detect variable objects using the Stetson J-index (Stetson 1996) which is specifically designed for programs
in which multiple observations are obtained closely spaced in time.  In a dataset consisting of pairs of magnitudes
with purely random noise, the J-index for a non-variable star tends toward zero; true variable objects will have positive J-index
values.  The residual of one magnitude measurement from the mean of all measurements is defined for each point on the lightcurve.
The residual values for each pair of observations are multiplied together and the J-index is the weighted sum of these combined
residuals over all pairs.  
The technique has a tendency to diminish the importance of individual bad data points (e.g. cosmic rays),
but the index will be more sensitive to signals in the data in which the pairs of observations deviate from the mean in the 
same direction.  Thus, the J-index should be more sensitive than the simple $\sigma$ clipping algorithm
described above to eclipsing variables, 
but it also amplifies defects in the data that affect both observations closely paired in time.  For example, 
magnitude variations due to blending with other objects is a function of the data quality which would be similar for
both individual measurements in the pair.  

\begin{figure}
\psfig{file=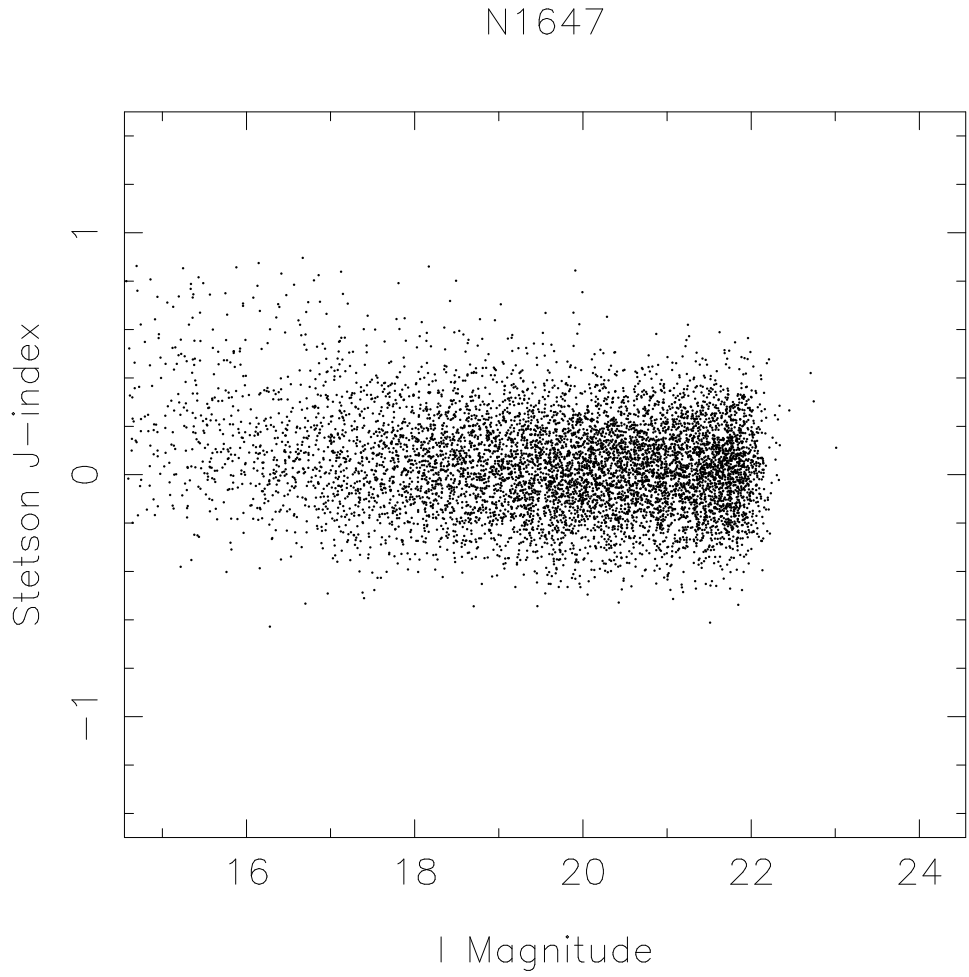,width=5truein,angle=-90}
\caption{The Stetson J-index for the sample of stars from the field of the NGC~1647 cluster.  Non-variable objects should tend to have
a J-index value of zero.  }
\label{fig:stetsonJ}
\end{figure}

\begin{figure}
\psfig{file=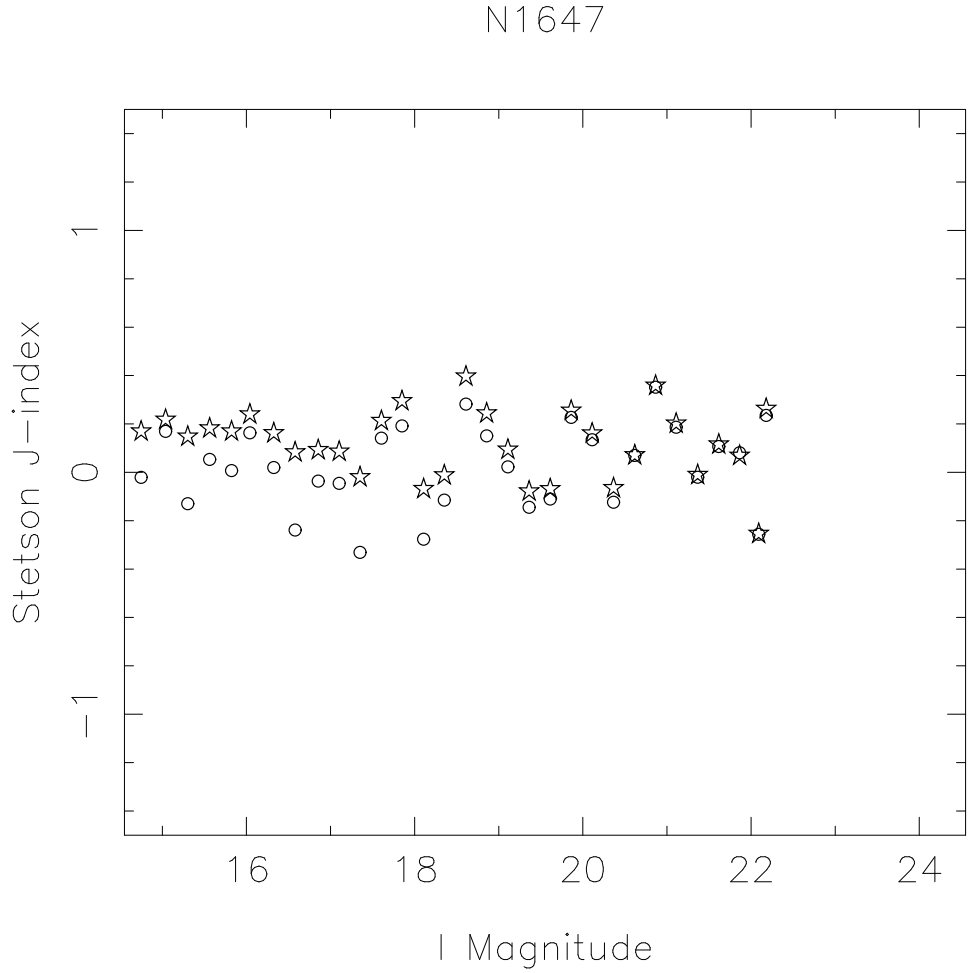,width=5truein,angle=-90}
\caption{The Stetson J-index for the sub-sample of stars from the field of the NGC~1647 cluster for which we artificially added an eclipse signal with amplitude of 0.05 magnitudes, period of 1.6 days, and duration of 2 hours.  
The symbols have the same meaning as in Figure~\ref{fig:rms_addeclipse}.  Open circles mark the original J-index value for the subset of stars; J-index values for the same stars with the added eclipse signal are marked by open stars. }
\label{fig:stetsonJ_addeclipse}
\end{figure}

Figure~\ref{fig:stetsonJ} shows a plot of the J-index as a function of 
I-band magnitude for objects in the field of NGC~1647.  
Like the rms plot, the J-index distribution is continuous, and there is no clear separation between variable and non-variable stars.  
Again, we add artificial eclipse signals to the same subset of stars discussed above and plot their original 
J-index values as large open circles in Figure~\ref{fig:stetsonJ_addeclipse} along with the J-index values of the lightcurves
with the added eclipse as large open stars.
The addition of the low amplitude eclipse increases the J-index value for all stars in the randomly chosen subset, but the eclipsing objects are still not clearly distinguishable from the majority of non-variable stars.  While we expected an improvment in the eclipse detection for the J-index over the simple $\sigma$-clipping, in fact this was not evident given the sampling of our data.
Neither technique described above is efficient at detecting the signatures of low-mass eclipsing binary stars in 
lightcurves with non-gaussian noise properties, thus we investigate algorithms designed more specifically for this purpose.
  
\subsection{Box-fitting Algorithm}
\label{sec:boxfilter}

Tingley (2003a,2003b) compared transit/eclipse detection algorithms and concluded that a modified version of the 
box fitting algorithm by Kov\'acs et~al. (2002, hereafter KZM) gave the best performance overall, although marginally, 
when compared to various matched filter and Bayesian algorithms.  Although no detection algorithm is clearly superior for
all eclipse signals of varying strengths, we chose to use the KZM algorithm 
since it is designed specifically to detect periodic signals which alternate between two
discrete levels, as in transits and eclipses.  The routine fits a
square step function to the phase-folded lightcurve: by assuming
the transiting system spends the majority of its time out of the
eclipse phase, this algorithm is able to find the best fitting function by minimizing
the simple expression $$ D= \sum_{n=1}^{N} \omega_n*x_n^2 - \frac{s^2}{r(1-r)}. $$

Here, $N$ is the total number of points in the lightcurve, 
$x_n$ is the value (differential magnitude) of each data point in the phase-folded lightcurve, $\omega_n$ is the
normalized weight of each point given by $\omega_n=\sigma_n^{-2}*(\sum_{i=0}^{N-1} \sigma_i^{-2})^{-1}$ (where $\sigma_i$ 
is the familiar gaussian standard deviation),  
$r$ is the sum of the weights during the eclipse/transit and $s$ is the weighted sum of $x_n$ over the in-transit points.  
The routine tests a grid of periods, transit phases, and durations, in order to find the minimum $D$.

Minimizing $D$ is equivalent to maximizing the Signal Residue, 
$$SR = Max[(\frac{s^2}{r*(1-r)})^{\frac{1}{2}}]$$ for each tested period.  Following KZM, we determine the
Signal Residue for each trial period, and finally calculate
the Signal Detection Efficiency, $$SDE = \frac{SR_{\rm peak} - <SR>}{\sigma_{SR}}$$ where $SR_{\rm peak}$ 
is the highest value of $SR$ over all tested periods, $<SR>$ is the 
the mean value of $SR$, and $\sigma_{SR}$ is the standard deviation.  
(We check by eye all potential eclipsing objects with $SDE \gtrsim 5$ depending on the sampling of the data.)
  
In our implementation of the algorithm we test a grid of periods and eclipse durations on all the lightcurves.  
We sample periods between 0.5 to 11 
days in intervals of 0.01 days for those clusters for which we have a sufficiently long temporal 
baseline of observations, as in M67, M35, and IC4756.  
For the other three clusters, we test periods to half the time extent of the observations.  
We test transit/eclipse durations between 0.02 and 0.25 days, $\sim 0.5$ to 6 hours.  

\begin{figure}
\psfig{file=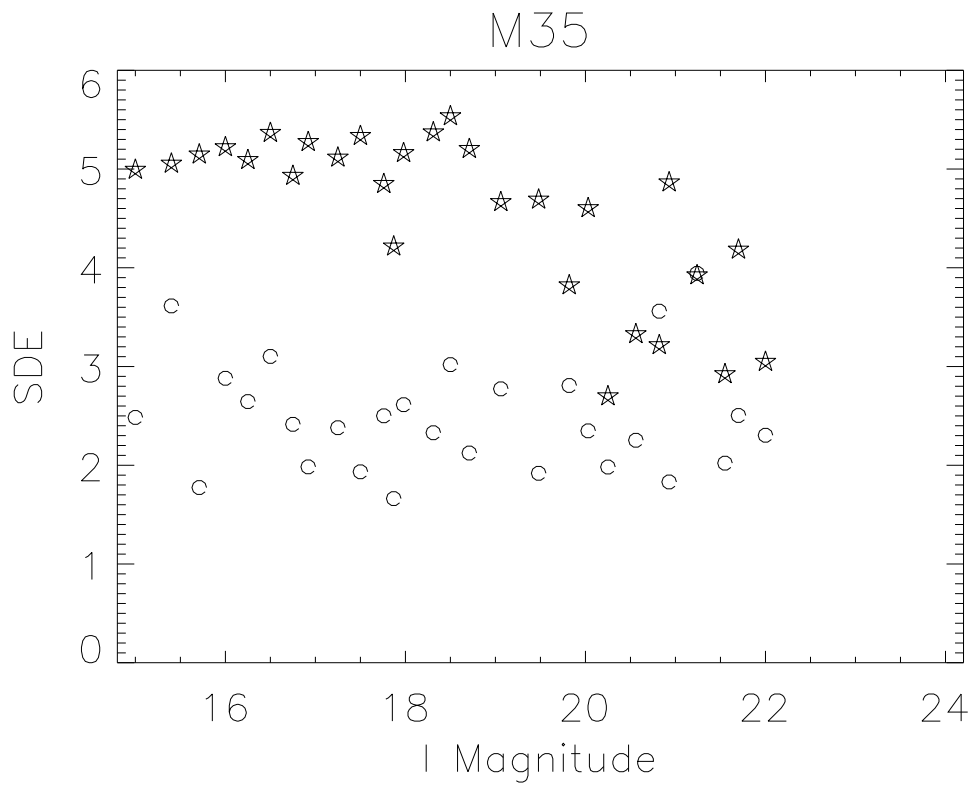,width=5truein,angle=0}
\caption{The signal detection efficiency (SDE) versus I-band magnitude output by the box-fitting algorithm 
for a set of simulated lightcurves with sampling frequency and rms values chosen to match the M35 cluster 
data.  We artificially added an eclipse signal with amplitude of 0.05 magnitudes, period of 1.6 days, and duration of 2 hours
to the simulated lightcurves.  The open circles mark the SDE value for the original set of lightcurves; 
SDE values for the lightcurves with the added eclipse signal are marked by the stars.  The algorithm recovers
the correct eclipse period for systems with $I\lesssim 20$ where the rms in the lightcurve is approximately equal to
the eclipse amplitude (0.05 magnitudes).
}
\label{fig:srperiod}
\end{figure}

The detection efficiency and the type of eclipsing variables which are most likely to be detected with this method depend on
the sampling function of the actual data and will be discussed in detail in further analysis papers.  Although in order
to illustrate the effectiveness of the algorithm, we 
show, in Figure~\ref{fig:srperiod}, a plot of the SDE 
versus I-band magnitude for a set of simulated lightcurves with rms values
and sampling frequency chosen to match the M35 cluster data.  We add an eclipse signal with a period of 1.6 days, a
duration of $\sim 2$ hours and an amplitude of 0.05 magnitudes to the lightcurves and compare the value of the period and SDE
recovered from the box-fitting algorithm.  The algorithm detects the eclipse signature in the lightcurves 
and recovers the correct period for stars with $I \lesssim 20$ 
where the rms of the lightcurve is approximately the eclipse amplitude (0.05 magnitudes).
Within this magnitude range, the simulated lightcurves with the eclipse signal have SDE values $\sim 5$ 
and are clearly distinguishable from the same lightcurves without the added eclipse.

\begin{figure}
\psfig{file=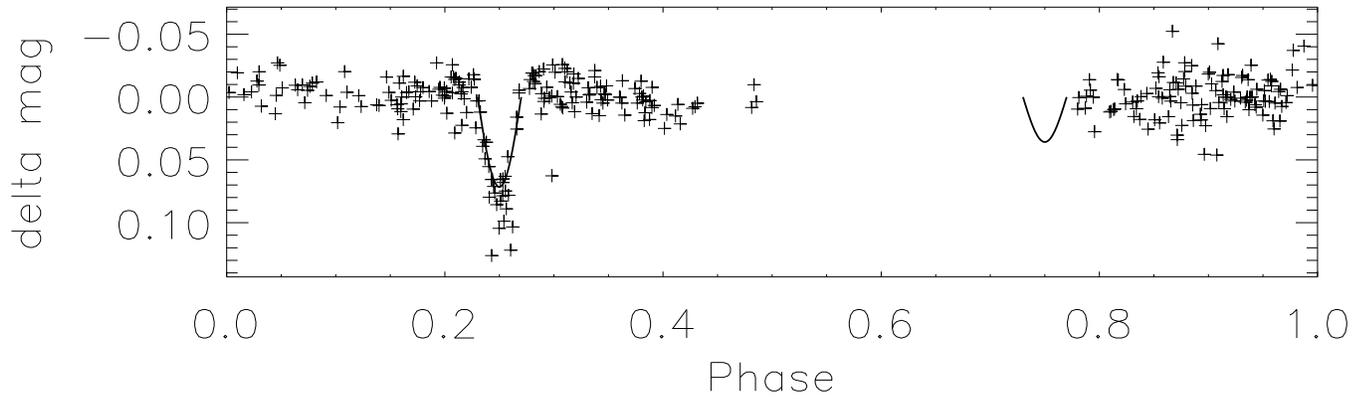,width=7truein,angle=90}
\caption{The phase-folded lightcurve for a candidate M-dwarf eclipsing binary star in the M35 cluster with a period of 1.09 days.  The eclipsing object was detected using the box-fitting algorithm described in \S~\ref{sec:boxfilter}.  The crosses show all differential photometry measurements for this object collected in January 2002, January 2003 and February 2003.  The solid line is a simple sine-wave fit to the measurements taken during the primary eclipse.  The phase of the secondary eclipse is marked by the same sine function fit to the primary eclipse, but with half the amplitude. }
\label{fig:m35lcurve}
\end{figure}

Further, in Figure~\ref{fig:m35lcurve}, we show an example of the phase-folded lightcurve of a candidate M-dwarf
eclipsing binary system in the M35 cluster detected using this technique.  
The best fitting period was found to be 1.09 days with an eclipse duration of $\sim $ 1 hour and amplitude of 0.07 magnitudes.

\section{Conclusions}
\label{sec:conclude}

We have described our survey of six galactic clusters designed to detect low mass eclipsing binary 
stars with known age and metallicity.  The data necessary to
calibrate and test stellar structure models of low-mass stars is still lacking, particularly mass and 
metallicity information on the same stars and sufficient numbers of accurate stellar radii 
measurments.  We explained the observations and processing steps applied to the survey data.  
We have also detailed our algorithms for generating lightcurves, selecting potential
cluster members, and determining potential eclipsing systems through their variability characteristics.
Calibrated photometry catalogues of all 
objects in the wide-FOV centered on each cluster, including all our variable object detections will follow in 
subsequent papers.  According to our simulations we expect to
find $\sim 5$ low-mass eclipsing binary stars within our clusters.  
We have processed the entire dataset including a determination of the 
photometric calibration equations for all photometric nights of data. 
We also plan use our data to define the degree of mass segregation present in each cluster which can be used to test dynamical evolution models of clusters and to study Galactic structure through analysis of the field populations in the data.

\acknowledgments

L.H. thanks Mike Irwin for helpful advice and for making available to us the CASU data processing software.  L.H. also thanks Imants Platais and Peter McCullough for discussions and suggestions.  L.H. is grateful support from the Metro Washington Chaper of the ARCS Foundation, Inc. and from the ZONTA International Amelia Earhart Fellowship Program.

\begin{table}
\caption{Open clusters}
\label{tab:cluslis}
\begin{tabular}{rcccccccccl}
\hline\noalign{\smallskip}
\hline\noalign{\smallskip}
\multicolumn{1}{r}{Cluster}    & 
\multicolumn{1}{c}{RA\tablenotemark{a} }  &
\multicolumn{1}{c}{DEC\tablenotemark{a} }  & 
\multicolumn{1}{c}{l} &
\multicolumn{1}{c}{b} &
\multicolumn{1}{c}{(M-m)$_{0}$} &  
\multicolumn{1}{c}{lg(age)} &
\multicolumn{1}{c}{$[\frac{Fe}{H}]$} &
\multicolumn{1}{c}{E(B-V)} &
\multicolumn{1}{c}{Radius\tablenotemark{b}} &
\multicolumn{1}{l}{Refs}
\\
\multicolumn{1}{r}{\phantom{Cluster}}    & 
\multicolumn{1}{c}{\phantom{$\alpha$}}  &
\multicolumn{1}{c}{\phantom{$\delta$}}  & 
\multicolumn{1}{c}{($^\circ$)}  &
\multicolumn{1}{c}{($^\circ$)}  & 
\multicolumn{1}{c}{\phantom{$(M-m)_{0}$}} &  
\multicolumn{1}{c}{\phantom{lg(age)}} &
\multicolumn{1}{c}{(dex)} &
\multicolumn{1}{c}{(mag)} &
\multicolumn{1}{c}{($^\prime$)} &
\multicolumn{1}{c}{\phantom{Refs}}
\\
\noalign{\smallskip}
\hline\noalign{\smallskip}
\noalign{\smallskip}
 NGC 1647      & 04 45.7   & +19 07 & 180&-17 & 8.7   & 8.3  & ...    & 0.37  	& 18	& 8,12 \\
 NGC 2168\tablenotemark{c} & 06 08.9 & +24 22 & 186&+2  & 9.8   & 8.2  & -0.21  & 0.2 & 20 & 1,2 \\
 NGC 2682\tablenotemark{d} & 08 51.5 & +11 49 & 215&+32 & 9.6  & 9.6  & -0.1   & 0.04 & 15 & 3,11,13 \\
 NGC 6633      & 18 27.5   & +06 34 & 36&+8  & 7.8  & 8.8  & -0.11  & 0.17  	&18	& 4,9,10,12 \\
 IC 4756       & 18 38.5   & +05 29 & 36&+5  & 7.6  & 8.9  & +0.04  & 0.23  	&26	& 4,5,10 \\
 NGC 6940      & 20 34.6   & +28 18 & 70&-7  & 9.8   & 9.0  & +0.04  & 0.2   	&32	& 5,6,7,14 \\
\tablenotetext{a}{J2000}
\tablenotetext{b}{The definitions of ``radius'' vary in the literature and are defined for the various clusters as follows:
N1647: this radius is the exponential scale length for stars of $m_{pg} \sim 15$ (12); 
N2168: there is a homogeneous distribution of cluster stars with $m/M_{\sun} < 1.6$ out to this radius (2);
N2682: surface density in cluster region $\sim$ background density at this radius for A5-G5 stars (13); 
N6633: this radius is the exponential scale length for stars of $m_{pg} \sim 15$ (12);
IC4756: this radius is the cluster nucleus for $M_V > 5$ stars (5); 
N6940: surface density in cluster region $\sim$ background density at this radius for $M_V > 10$ (14).
}
\tablenotetext{c}{M35}
\tablenotetext{d}{M67}
\tablenotetext{}{REFERENCES--(1)Sung \& Bessell 1999;(2)Barrado~y~Navascu\'es et~al. 2001;
(3)Fan et~al. 1996;(4)Robichon et~al. 1999;(5)Lynga 1987;(6)Laarson-Leander 1964;
(7)Van den Berg, M. \& Verbunt 2001;(8)Turner 1992;(9)Jeffries et~al. 2000;(10)Briggs et~al. 2000;(11)Sandquist 2004;
(12) Francic 1989;(13)Bonatto \& Bica 2003.;(14)Nilakshi et al. (2002)}
\end{tabular}
\end{table}

\clearpage

\begin{table}
\caption{Observations.}
\label{tab:obs}
\begin{tabular}{rrrrcrc}
\hline\noalign{\smallskip}
\hline\noalign{\smallskip}
\multicolumn{1}{c}{Cluster} &
\multicolumn{1}{c}{Filter} &
\multicolumn{1}{c}{Observing} &
\multicolumn{1}{c}{Site} &
\multicolumn{1}{c}{Exposure \tablenotemark{a}} &
\multicolumn{1}{c}{No. \tablenotemark{b}} &
\multicolumn{1}{c}{Mean sampling} 
\\
\multicolumn{1}{c}{\phantom{Cluster}} &
\multicolumn{1}{c}{\phantom{Filter}} &
\multicolumn{1}{c}{dates} &
\multicolumn{1}{c}{\phantom{Site}} &
\multicolumn{1}{c}{time (sec)} &
\multicolumn{1}{c}{frames} &
\multicolumn{1}{c}{rate($hr^{-1}$)} 
\\
\noalign{\smallskip}
\hline\noalign{\smallskip}
\noalign{\smallskip}
          NGC 2168 & Cousins I          &         25-28 Jan 03  &  KPNO   & 3 \& 50      & 46 \& 133   & $\sim$0.8 \\
 \phantom{NGC 2168}& Cousins I          &          8-10 Feb 03  &  INT    & 7 \& 50      & 126 \& 244  & $\sim$0.2  \\
 \phantom{NGC 2168}& Cousins I          &          9-10 Feb 03  &  KPNO   & 3 \& 50      & 67 \& 133   & $\sim$0.4 \\
 \phantom{NGC 2168}& Cousins I          &          3-5 Feb 02  &   INT    & 20 \& 90     & 20 \& 70 &  $\sim$0.7 \\
 \phantom{NGC 2168}& Johnson V          &          25-28 Jan 03  &  KPNO  & 5 \& 150     & 13 \& 52 &  $\sim$1.9 \\
 \phantom{NGC 2168}& Johnson V          &          3-5 Feb 02  &    INT   & 60 \& 300    & 20 \& 72 &  $\sim$0.7 \\
          NGC 2682 & Cousins I          &         25-28 Jan 03  &  KPNO   & 3 \& 40      & 34 \& 112   & $\sim$0.8 \\
 \phantom{NGC 2682}& Cousins I          &          8-10 Feb 03  &  INT    & 7 \& 50      & 66 \& 132   & $\sim$0.2  \\
 \phantom{NGC 2682}& Cousins I          &          9-10 Feb 03  &  KPNO   & 3 \& 40      & 15 \& 32    & $\sim$0.4\\
 \phantom{NGC 2682}& Cousins I          &          3-5 Feb 02  &    INT   & 20 \& 90     & 13 \& 38 &  $\sim$1.3 \\
 \phantom{NGC 2682}& Johnson V          &         25-28 Jan 03  &  KPNO   & 5 \& 100     & 12 \& 48 &  $\sim$1.6\\
 \phantom{NGC 2682}& Johnson V          &          3-5 Feb 02  &    INT   & 60 \& 300    & 10 \& 36 &  $\sim$1.3 \\
          NGC 6940 & Cousins I          &         12-17 Sep 02  &  KPNO   & 1 \& 50      & 85 \& 177   & $\sim$0.6 \\
 \phantom{NGC 6940}& Johnson V          &         12-17 Sep 02  &  KPNO   & 1 \& 100     & 19 \& 55 &  $\sim$1.6\\
          NGC 1647 & Cousins I          &         12-17 Sep 02  &  KPNO   & 1 \& 20      & 46 \& 102   & $\sim$0.8 \\
 \phantom{NGC 1647}& Johnson V          &        12-17 Sep 02  &   KPNO   & 1 \& 60      & 12 \& 27 &  $\sim$1.3\\
          NGC 6633 & Cousins I          &           3-7 Jul 02  &  INT    & 3 \& 20      & 216 \& 653  & $\sim$0.4 \\
 \phantom{NGC 6633}& Johnson V          &           3-7 Jul 02  &  INT    & 10 \& 75     & 51 \& 102 & $\sim$0.9\\
           IC 4756 & Gunn i             &         19-22 Jun 03  &  INT    & 3 \& 30      & 203 \& 404  & $\sim$0.3 \\
  \phantom{IC 4756}& Cousins I		&      24-27 Jun 03  &     KPNO   & 3 \& 50       & 56 \& 118 &  $\sim$0.4\\ 
  \phantom{IC 4756}& Johnson V		&      19-22 Jun 03  &      INT   & 3 \& 100      & 42 \& 84 &  $\sim$0.6\\ 
  \phantom{IC 4756}& Johnson V		&      24-27 Jun 03  &     KPNO   & 3 \& 150      & 18 \& 31 &  $\sim$1.6\\ 
\tablenotetext{a} {Both the short and long exposure times are listed. }
\tablenotetext{b} {The number of exposures taken for each exposure time listed.}
\end{tabular}
\end{table}

\clearpage

\begin{table}
\caption{KPNO color terms }
\label{tab:cterms}
\begin{tabular}{rccccccccc}
\hline\noalign{\smallskip}
\hline\noalign{\smallskip}
\multicolumn{1}{c}{Obs Run}    &
\multicolumn{1}{c}{filter}    &
\multicolumn{1}{c}{chip1}    &
\multicolumn{1}{c}{chip2}  &
\multicolumn{1}{c}{chip3}  &
\multicolumn{1}{c}{chip4} &
\multicolumn{1}{c}{chip5} &
\multicolumn{1}{c}{chip6} &
\multicolumn{1}{c}{chip7} &
\multicolumn{1}{c}{chip8}
\\
\noalign{\smallskip}
\hline\noalign{\smallskip}
\noalign{\smallskip}
 Sep 2002 & V & -0.018 & -0.014 & -0.026 & -0.024 & -0.022 & -0.015  & -0.023 & -0.022   \\
 Jun 2003 & V & -0.018 & -0.017  & -0.020 & -0.030  & -0.026 & -0.019  & -0.016 & ...   \\
 Average &  V & -0.018 & -0.016 & -0.023 & -0.027 & -0.024 & -0.017 & -0.020& -0.022 \\
 \\
 Sep 2002 & I & 0.024 & -0.003 & 0.019 & 0.021   & 0.022 & 0.020   & 0.011 & 0.020 \\
 Jun 2003 & I & 0.038 & 0.037  & 0.029 & 0.017  & 0.028 & 0.023  & 0.015 & ... \\
 Average &  I & 0.031 & 0.017 & 0.024 & 0.019 & 0.025 & 0.022 & 0.013 & 0.020 \\
\end{tabular}
\end{table}

\clearpage 

\begin{table}
\caption{KPNO extinction terms and zero points for photometric nights}
\label{tab:extterms}
\begin{tabular}{ccccc}
\hline\noalign{\smallskip}
\hline\noalign{\smallskip}
\multicolumn{1}{c}{Date} &
\multicolumn{1}{c}{V zero point} &
\multicolumn{1}{c}{I zero point} &
\multicolumn{1}{c}{V extinction term} &
\multicolumn{1}{c}{I extinction term}
\\
\noalign{\smallskip}
\hline\noalign{\smallskip}
\noalign{\smallskip}
12 Sep 2002 &$ 24.95\pm 0.02$ &$ 24.55\pm 0.01$  &$-0.083\pm0.006$ &$ -0.064\pm0.003$  \\
13 Sep 2002 &$ 24.93\pm 0.02$ &$ 24.54\pm 0.01$  &$-0.097\pm0.016$ &$ -0.081\pm0.001$ \\
14 Sep 2002 &$ 24.97\pm 0.02$ &$ 24.57\pm 0.01$  &$-0.134\pm0.004$ &$ -0.064\pm0.002$ \\
15 Sep 2002 &$ 24.95\pm 0.02$ &$ 24.56\pm 0.01$  &$-0.174\pm0.009$ &$ -0.121\pm0.003$ \\
16 Sep 2002 &$ 24.94\pm 0.02$ &$ 24.55\pm 0.01$  &$-0.150\pm0.010$ &$ -0.169\pm0.003$ \\
25 Jan 2003 & $24.99 \pm 0.02$ & $24.59 \pm 0.04$ & $-0.152 \pm  0.143$  & $-0.042 \pm  0.018$  \\
26 Jan 2003 & $24.99 \pm 0.01$ & $24.59 \pm 0.05$ & $-0.135 \pm  0.002$  & $-0.057 \pm  0.030$  \\
27 Jan 2003 & $24.97 \pm 0.01$ & $24.59 \pm 0.01$ & $-0.124 \pm  0.017$  & $-0.029 \pm  0.022$  \\
28 Jan 2003 & $24.97 \pm 0.02$ & $24.58 \pm 0.04$ & $-0.124 \pm  0.029$  & $-0.039 \pm  0.026$  \\
\end{tabular}
\end{table}

\clearpage

\clearpage

\clearpage

\clearpage

\clearpage

\clearpage

\clearpage

\clearpage

\clearpage


\clearpage

\clearpage

\clearpage

\clearpage

\clearpage

\clearpage

\clearpage

\end{document}